\documentclass[journal]{IEEEtran}
\usepackage{amsmath,amsfonts,amssymb}

\usepackage{graphicx}
\usepackage{pgfplots}
\usepackage{booktabs}
\pgfplotsset{compat=1.18}
\usepackage{circuitikz}
\allowdisplaybreaks
\usetikzlibrary{arrows.meta,positioning}

\usepackage{changepage}

\usepackage{diagbox}

\usepackage{balance}

\usetikzlibrary{decorations.markings} 
\usepackage[caption=false,font=normalsize,labelfont=sf,textfont=sf]{subfig}

\usepackage{textcomp}
\usepackage{stfloats}
\usepackage{url}
\usepackage{verbatim}
\usepackage{array}
\usepackage{makecell}
\usepackage{cite} 

\usepackage{tikz}

\usepackage{cuted}      
\usepackage{lipsum}     
\usepackage{stfloats}    
\usepackage{amsmath}    
\usepackage{adjustbox}

\usepackage{xcolor}
\usepackage{bm}
\usepackage{upgreek}
\usepackage{hyperref}
\usepackage{textcomp} 
\usepackage{multirow}
\usepackage{fix-cm}    
\usepackage[table]{xcolor}
\usepackage{calc}

\def\BibTeX{{\rm B\kern-.05em{\sc i\kern-.025em b}\kern-.08em
    T\kern-.1667em\lower.7ex\hbox{E}\kern-.125emX}}
\begin{document}

\title{Shadow Pricing of Static Voltage Stability Services within Unit Commitment for Inverter-Dominated Power Systems}
\author{Peng Wang, \IEEEmembership{Student Member, IEEE}, Luis Badesa
\thanks{This work was supported by MICIU/AEI/10.13039/501100011033 and ERDF/EU under grant PID2023-150401OA-C22, as well as by the Madrid Government (Comunidad de Madrid-Spain) under the Multiannual Agreement 2023-2026 with Universidad Politécnica de Madrid, `Line A - Emerging PIs' (grant number: 24-DWGG5L-33-SMHGZ1). The work of Peng Wang was also supported by China Scholarship Council under grant 202408500065.}
\thanks{Peng Wang and Luis Badesa are with the School of Industrial Engineering and Design (ETSIDI), Technical University of Madrid (UPM), Spain (e-mail: peng.wang@alumnos.upm.es, luis.badesa@upm.es). \textit{Corresponding author: Luis Badesa.}}
}

\maketitle

\begin{abstract}
Modern power systems are increasingly dominated by Inverter-Based Resources (IBR), most of which work in Grid-following (GFL) mode. This implies that they do not directly control their terminal voltage, so the static voltage stability at these buses may  be compromised, especially under constant-power-factor operation that lacks voltage-adaptive reactive support. In addition, weather-driven IBR are often installed in electrically remote areas with low Short-Circuit Ratio (SCR), further exacerbating voltage issues. To address this challenge, grid-forming control can be utilized to enhance low-SCR buses, while GFL-IBR could be explicitly required to provide voltage support through grid codes. As an alternative, a market mechanism could be devised that incentivizes relevant generators to proactively adjust their operating points as a service to maintain voltage stability, while the theoretical framework for such a market has not been developed. To fill this gap, this work adopts a second-order cone-based static voltage stability constraint for GFL-IBR buses within a unit commitment problem, and proposes a mechanism to assign shadow prices to this ancillary service. To determine appropriate price values under non‑convex conditions, different pricing schemes are assessed. Using a modified IEEE 30-bus system, we demonstrate that both the dispatchable and restricted pricing methods can yield revenue-adequate service prices, though the former may deliver less efficient price signals and the latter may require well-defined uplift payments. This implies that, given differentiated pricing mechanisms and price signals, operators need to select a suitable pricing method in accordance with actual system conditions and market rules.
\end{abstract}


\vspace{-0.3cm}
\section*{Nomenclature}
\addcontentsline{toc}{section}{Nomenclature}

\subsection*{Indices and Sets}
\begin{IEEEdescription}
    \item[$g_c,\mathcal{G}_c$] Index, Set of conventional SGs
    \item[$g_f, \mathcal{G}_f$] Index, Set of GFL-IBR  
    \item[$g_v,\mathcal{G}_v$] Index, Set of VSGs
    \item[$i,\mathcal{I}$] Index, Set of buses 
    \item[$m,\mathcal{M}$] Index, Set for products of units' operating states
    \item[$\Phi(\cdot)$] Index of generator buses
\end{IEEEdescription}

\subsection*{Constants and Parameters}
\begin{IEEEdescription}
    \item[$\mathrm{c}_{g_c}^\mathrm{m}$] \qquad Marginal generation cost of SG (\texteuro/MWh)
    \item[$\mathrm{c}_{g_c}^\mathrm{nl}$] \qquad No-load cost of SG (\texteuro/h)
    \item[$\mathrm{c}^\mathrm{st}_{g_c},\mathrm{c}^\mathrm{sh}_{g_c}$] \qquad Startup/shutdown costs of SGs (\texteuro) 
    \item[$\mathrm{P}_i^{\mathrm{D}}, \mathrm{Q}_i^{\mathrm{D}}$] \qquad \parbox[t]{.9\linewidth} {Active/reactive demand at bus $i$ (MW, Mvar)}
    \item[$\mathrm{P}_{\{\cdot\}}^{\mathrm{min}}, \mathrm{P}_{\{\cdot\}}^{\mathrm{max}}$] \qquad  \parbox[t]{.9\linewidth} {Generators' active power output limits (MW)}
    \item[$ \mathrm{Q}_{\{\cdot\}}^{\mathrm{min}}, \mathrm{Q}_{\{\cdot\}}^{\mathrm{max}}$] \qquad  \parbox[t]{.9\linewidth} {Generators' reactive power output limits (Mvar)}
    \item[$ \mathrm{S}^{\mathrm{max}}_{ \{g_c,g_f,g_v\} }$] \qquad Rated apparent power of generators (MVA)
    \item[$\alpha_{g_f},\alpha_{g_v}$] \qquad \parbox[t]{.9\linewidth} {Capacity percentage of GFL-IBR and VSGs}
\end{IEEEdescription}

\subsection*{Primal and Dual Variables}
\begin{IEEEdescription}
    \item[$C_{g_c}^\mathrm{st},C_{g_c}^\mathrm{sh}$] \qquad \parbox[t]{.9\linewidth} {Costs incurred by startup/shutdown of SG (\texteuro)}
    \item[$\hat{P}_{g_f}, \hat{Q}_{g_f}$] \qquad \parbox[t]{.9\linewidth} {Equivalent active/reactive power injected by GFL-IBR (MW, Mvar)}
    \item[$P_{\{\cdot\}}, Q_{\{\cdot\}}$] \qquad \parbox[t]{.9\linewidth} {Active/reactive power output of generators (MW, Mvar)}
    \item[$u_{g_c}$] \qquad Binary variable, commitment of SG
    \item[$\Gamma_{g_f}$] \qquad Variable associated with SCR 
    \item[$\eta_m$] \qquad Product of operating states of each two (V)SGs
    \item[$\lambda_{1,g_f},\lambda_{2,g_f},\mu_{g_f}$] \qquad \qquad \parbox[t]{.8\linewidth} {Dual variables, associated with voltage stability constraints}
    \item[$\lambda^\mathrm{E}$] \qquad Dual variable, energy price (\texteuro/MWh)
    \item[$\lambda_{g_c,\mathrm{commit}}$] \qquad Dual variable, commitment price for SG (\texteuro)
\end{IEEEdescription}

\section{Introduction}
\label{sec:introduction}
\IEEEPARstart{V}{oltage} stability analysis conventionally focuses on load buses, under the assumption that voltages at synchronous-type controlled buses are properly regulated \cite{chu2022voltage}. As a result, voltage stability is largely affected by system loading conditions. However, as Grid-Following (GFL) Inverter-Based Resources (IBR) become increasingly prevalent in low-carbon power systems, voltage stability at GFL-IBR buses must also be considered to ensure secure system operation.

This consideration arises from the fact that, unlike Grid-Forming (GFM) generators, which behave as voltage sources and establish system voltages, GFL-IBR interfaced through power electronics typically operate as controlled current sources synchronized to the grid. Consequently, these IBR do not directly regulate terminal voltage and generally do not provide reactive power support to maintain voltage stability unless explicitly required by grid codes. In addition, due to the spatial distribution of Renewable Energy Sources (RES), such as wind and solar power, IBR with inherently low short-circuit capacity are often installed in remote areas electrically distant from the main grid. This geographical separation further weakens the Short-Circuit Ratio (SCR) at these locations and increases the risk of voltage instability \cite{hosseinzadeh2021voltage}. In particular, interactions among multiple IBR may become detrimental when they are electrically close to one another \cite{chu2022voltage,qays2023system}. Therefore, it is important to incorporate this phenomenon into voltage stability analysis at GFL-IBR buses as well.

A wide range of approaches has been proposed to enhance voltage stability. References \cite{savvopoulos2019long,roy2013reactive} investigate planning strategies for reactive support resources (e.g., static synchronous compensators, capacitors and reactors) to improve voltage stability. For regions with existing assets, \cite{wu2025enhancing} proposes converting retired thermal units into synchronous condensers to reinforce system strength, while \cite{guo2022control} mitigates instabilities by configuring energy storage converters as voltage-controlled inverters. However, these measures will bring additional investment costs to system operators. Dynamic control methods have also attracted considerable attention, with \cite{khan2024grid} advocating for a transition of IBR control from GFL to GFM modes to actively strengthen voltage stability. Yet this approach requires well-coordinated mode switching, as abrupt transitions may otherwise interfere with converter control loops and undermine system stability \cite{kim2025seamless}. From a perspective of system dispatch, \cite{mandoulidis2022overview} integrates load shedding and reactive support from IBR into voltage stability assessment. Similarly, \cite{chu2022voltage} demonstrates how coordinated active and reactive power injections from GFL-IBR can contribute to voltage stability enhancement.

Although technical measures, such as optimizing investments in reactive support technologies \cite{savvopoulos2019long,roy2013reactive}, improving the utilization of system components \cite{wu2025enhancing,guo2022control}, and refining IBR control strategies \cite{khan2024grid,kim2025seamless}, have proven effective in securing voltage stability, they do not explicitly consider static voltage stability at GFL-IBR buses during system scheduling, where system-wide resources can be utilized to maintain voltage stability. Specifically, conventional Synchronous Generators (SGs) inherently introduce subtransient reactance in parallel with the existing network impedance, thereby directly reducing the equivalent impedance seen from the bus. Virtual Synchronous Generators (VSGs) can achieve a similar effect by actively emulating a voltage source with adjustable virtual impedance \cite{du2026improved}, which also enhances SCR without requiring the installation of extra synchronous machines. Meanwhile, GFL-IBR can be required to properly coordinate their active and reactive injections to support local voltage stability \cite{chu2022voltage}.

Based on the characteristics of different resources analyzed above, one way to address the voltage stability challenge at GFL-IBR buses could be to enforce grid codes that mandate eligible generators to provide such static voltage stability support. However, it is worth exploring a potentially more cost-effective alternative: a market mechanism that remunerates market participants for providing ancillary services only when system requires them. The goal of such a mechanism would be to create appropriate incentives for different generators to provide only the right amount of the voltage stability service at times when they are really needed, therefore avoiding over-investment in IBR assets that could lead to increased costs for relevant entities. This measure is also supported by \cite{banshwar2017renewable}, which suggests incorporating voltage stability as an ancillary service in electricity markets.

Previous studies have sought to create financial incentives for generators to maintain voltage stability. The authors in \cite{chattopadhyay2003spot,kim2011market,potter2023reactive} derive nodal active and reactive prices from the dual variables of optimal power flow, thereby incentivizing generators to support voltage stability. Reference \cite{chung2004cost} links the voltage stability margin to reactive power costs in the objective function and derives service prices through cost allocation principles. These studies provide valuable insights into evaluating the economic value of voltage stability services.

Nevertheless, two critical points of service pricing remain insufficiently explored. The first concerns the binary nature of Unit Commitment (UC), while the second relates to the pricing of SCR enhancement, which has already been incorporated into ancillary services by the Australian Energy Market Operator (AEMO) \cite{gu2019review}. The calculation of SCR is closely related to the system equivalent impedance, which incorporates the self-impedance of SGs. Since the complete contribution of SGs' self-impedance to the network depends on generator commitment status, explicit consideration of UC is necessary for accurately tracking system impedance dynamics. Accordingly, joint consideration of UC, reactive power support, and SCR enhancement is required in voltage stability service pricing to enable efficient coordination and utilization of diverse generation resources.

In this context, to fill the gap identified above, this article employs a Second-Order Cone (SOC) formulation proposed in \cite{chu2022voltage} to represent how static voltage stability constraints on GFL-IBR buses relate to generators through power injections and SCR. Taking that model as a starting point, a shadow-pricing formulation for the voltage stability as an ancillary service is proposed, enabling generators that support this constraint to receive financial compensation. However, incorporating UC into the pricing framework introduces a major challenge arising from the non-convexity caused by binary commitment variables, which precludes the direct derivation of Karush–Kuhn–Tucker (KKT) conditions and consequently hinders the formation of market clearing prices and ancillary service remuneration.  

Consequently, dedicated pricing approaches are required. To determine shadow prices associated with least-cost UC and dispatch under static voltage stability constraints, this paper investigates three pricing models: dispatchable pricing, restricted pricing \cite{gribik2007market}, and marginal unit pricing \cite{chu2024pricing}. Their implementation mechanisms and inherent limitations in market outcomes are thoroughly analyzed in this paper.

The contributions of this work are as follows:
\begin{enumerate}
    \item To decompose the dual variables of the existing static voltage stability constraint into two distinct ancillary service components, namely SCR enhancement from (V)SGs and reactive power support from GFL-IBR, thereby justifying that active power should not be remunerated for voltage stability.
    \item To deliver new insights on the incentives for generators to offer ancillary services under the proposed pricing framework, and the implications of different methodologies for dealing with non-convexities.
    \item To demonstrate that both proposed pricing methods can provide adequate financial incentives to generating units: the dispatchable method, in which relaxed commitment variables may lead to less efficient price signals, and the restricted method, which can yield accurate prices but may require appropriate uplift payments for the associated units.
\end{enumerate}

The remainder of this paper is structured as follows. Section~\ref{Methodologies for Shadow Prices} models static voltage stability constraints and derives shadow prices for voltage stability services. Section~\ref{Pricing Voltage Stability Services} introduces pricing schemes for computing service prices under a voltage stability constrained UC model. Section~\ref{Case Studies} presents case studies to examine the proposed pricing methodologies. Section~\ref{Conclusion} concludes the paper and discusses future work.

\section{Assigning Shadow Prices to Static Voltage Stability Services}\label{Methodologies for Shadow Prices}
This section begins by introducing the voltage stability constraint for GFL-IBR buses. It then describes an offline training to approximate the nonlinear terms in this constraint. Finally, dual variables are assigned to the constraint to derive shadow prices for voltage stability services. It is noteworthy that the development of pricing mechanism in Section~\ref{Shadow Prices for Voltage Stability Constraint} is built upon the SOC constraint structure from \cite{chu2022voltage}.

\subsection{Representation of Static Voltage Stability Constraint}\label{Representation of Static Voltage Stability Constraint}

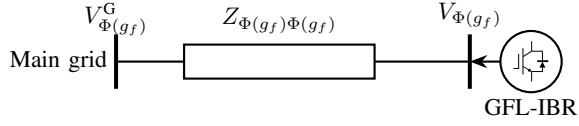
\begin{figure}[!t]
\centering
\begin{tikzpicture}[>=Stealth, thick, scale=0.9, transform shape]
\draw [line width=1.5pt] (-3,0.4) -- (-3,-0.4);
\node[left] at (-3,0) {Main grid};  
\node[below] at (-3,1) {$V^\mathrm{G}_{\Phi(g_f)}$};  

\draw (-3,0) -- (-2,0);  
\draw (-2,0.25) rectangle (0.8,-0.25);  
\node at (-0.6,0.6) {$Z_{\Phi(g_f)\Phi(g_f)}$};  
\draw (0.8,0) -- (2.2,0);  

\draw [line width=1.5pt] (2.2,0.4) -- (2.2,-0.4);  
\node[below] at (2.2,1) {$V_{\Phi(g_f)}$};  

\draw[->, thick] (3.1-0.45,0) -- (2.2,0);

\draw (3.1,0) circle (0.45);

\node at (3.1,0) {
  \includegraphics[
    width=0.65cm,    
    height=0.65cm,   
    keepaspectratio, 
    clip             
  ]{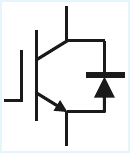}  
};

\node[below] at (3.1,-0.42) {GFL-IBR};

\end{tikzpicture}
         \vspace{-0.3cm}
\caption{Equivalent circuit of the grid seen from a GFL-IBR bus.}
\label{Z_define}
\vspace{-0.4cm}
\end{figure}

At steady state, considering interactions among different GFL-IBR, the voltage stability condition derived from the equivalent two-bus system in Fig.~\ref{Z_define} is formulated as follows:
\begin{subequations}\label{eq:explanation_voltage_stability}
\begin{align}
& \hat{P}_{g_f}^{2}+\hat{Q}_{g_f}^{2} \leq \big( \hat{Q}_{g_f} + \frac{|V^\mathrm{G}_{\Phi(g_f)}|^2}{2|Z_{\Phi(g_f)\Phi(g_f)}|}  \big)^{2} \\
& \hat{P}_{g_f}=P_{g_f}+ \hspace{-0.5cm} \sum_{g_f^{\prime}\in\mathcal{G}_f,g_f^{\prime}\neq g_f}\frac{|Z_{\Phi(g_f)\Phi(g_f^{\prime})}|}{|Z_{\Phi(g_f)\Phi(g_f)}|} \frac{V_{\Phi(g_f)}}{V_{\Phi(g_f')}} P_{g_f^{\prime}}  \\
& \hat{Q}_{g_f}=Q_{g_f}+ \hspace{-0.5cm} \sum_{g_f^{\prime}\in\mathcal{G}_f,g_f^{\prime}\neq g_f}\frac{|Z_{\Phi(g_f)\Phi(g_f^{\prime})}|}{|Z_{\Phi(g_f)\Phi(g_f)}|}  \frac{V_{\Phi(g_f)}}{V_{\Phi(g_f')}} Q_{g_f^{\prime}} 
    \end{align}
\end{subequations}
where $V^\mathrm{G}_{\Phi(g_f)}$ and $V_{\Phi(g_f)}$ denote the voltages at the equivalent grid bus and the GFL-IBR bus, respectively. $Z_{\Phi(g_f)\Phi(g_f)}$ denotes the impedance between the two buses. The equivalent active/reactive power injections of IBR `$g_f$', i.e., $\hat{P}_{g_f}$ and $\hat{Q}_{g_f}$, account for both local injections ($P_{g_f}$ or $Q_{g_f}$) and injections from other IBR `$g_f'$' ($P_{g_f'}$ or $Q_{g_f'}$), referred to their locations via the impedance ratio and voltage. The constraints \eqref{eq:explanation_voltage_stability} show that the admissible active power injection from the IBR is bounded by a limit determined by reactive power injection and the short-circuit capacity at bus $\Phi(g_f)$.

Given the assumptions that $V_{\Phi(g_f)}/V_{\Phi(g_f')} \approx 1$ and $|V^\mathrm{G}_{\Phi(g_f)}|^2 \approx 1$ hold under normal operating conditions in power systems \cite{chu2022voltage}, the static voltage stability condition can be further expressed as:
\begin{subequations}\label{eq:define_of_original_VS_constraints}
\begin{align}
& \hat{P}_{g_f}^{2}+\hat{Q}_{g_f}^{2} \leq \big( \hat{Q}_{g_f} + \Gamma_{g_f}  \big)^{2} \label{eq:define_VS_full} \\
& \hat{P}_{g_f}=P_{g_f}+ \hspace{-0.5cm} \sum_{g_f^{\prime}\in\mathcal{G}_f,g_f^{\prime}\neq {g_f}}\frac{|Z_{\Phi({g_f})\Phi(g_f^{\prime})}|}{|Z_{\Phi({g_f})\Phi({g_f})}|}P_{g_f^{\prime}} \label{eq:define_VS_P} \\
& \hat{Q}_{g_f}=Q_{g_f}+ \hspace{-0.5cm} \sum_{g_f^{\prime}\in\mathcal{G}_f,g_f^{\prime}\neq {g_f}}\frac{|Z_{\Phi({g_f})\Phi(g_f^{\prime})}|}{|Z_{\Phi({g_f})\Phi({g_f})}|}Q_{g_f^{\prime}}   \label{eq:define_VS_Q} \\
& \Gamma_{g_f}=\frac{1}{2|Z_{\Phi({g_f})\Phi({g_f})}|}  \label{eq:define_Gamma}
    \end{align}
\end{subequations}
where $\Gamma_{g_f}$ is associated with the SCR of IBR bus $\Phi(g_f)$, forming a relationship:
\begin{equation}\label{eq:SCR}
    \mathrm{SCR}_{\Phi(g_f)}=2\Gamma_{g_f}=\frac{1}{|Z_{\Phi({g_f})\Phi({g_f})}|}
\end{equation}

\begin{figure}[!t]
    \centering
\begin{tikzpicture}
\begin{axis}[
    width=9cm,
    height=3.5cm,
    ymajorgrids=true,          
    grid style={dashed},
    xlabel={\scriptsize $\hat Q_{g_f}\ \mathrm{[p.u.]}$},
    ylabel={\scriptsize $\hat P_{g_f}\ \mathrm{[p.u.]}$},
    xmin=-1, xmax=1,
    ymin=0, ymax=2,
    label style={font=\scriptsize}, 
    tick label style={font=\scriptsize},
    xtick={-1,-0.5,0,0.5,1},
    samples=200, 
    smooth
]

\draw[red!60, thin] (axis cs:-1.00, 0.0) -- (axis cs:-1.00, 2.0);
\draw[red!60, thin] (axis cs:-0.96, 0.0) -- (axis cs:-0.96, 2.0);
\draw[red!60, thin] (axis cs:-0.92, 0.0) -- (axis cs:-0.92, 2.0);
\draw[red!60, thin] (axis cs:-0.88, 0.0) -- (axis cs:-0.88, 2.0);
\draw[red!60, thin] (axis cs:-0.84, 0.0) -- (axis cs:-0.84, 2.0);
\draw[red!60, thin] (axis cs:-0.80, 0.0) -- (axis cs:-0.80, 2.0);

\draw[red!60, thin] (axis cs:-0.76, 0.0) -- (axis cs:-0.76, 2.0);
\draw[red!60, thin] (axis cs:-0.72, {sqrt(3*(-0.72)+2.25)}) -- (axis cs:-0.72, 2.0);
\draw[red!60, thin] (axis cs:-0.68, {sqrt(3*(-0.68)+2.25)}) -- (axis cs:-0.68, 2.0);
\draw[red!60, thin] (axis cs:-0.64, {sqrt(3*(-0.64)+2.25)}) -- (axis cs:-0.64, 2.0);
\draw[red!60, thin] (axis cs:-0.60, {sqrt(3*(-0.60)+2.25)}) -- (axis cs:-0.60, 2.0);
\draw[red!60, thin] (axis cs:-0.56, {sqrt(3*(-0.56)+2.25)}) -- (axis cs:-0.56, 2.0);
\draw[red!60, thin] (axis cs:-0.52, {sqrt(3*(-0.52)+2.25)}) -- (axis cs:-0.52, 2.0);
\draw[red!60, thin] (axis cs:-0.48, {sqrt(3*(-0.48)+2.25)}) -- (axis cs:-0.48, 2.0);
\draw[red!60, thin] (axis cs:-0.44, {sqrt(3*(-0.44)+2.25)}) -- (axis cs:-0.44, 2.0);
\draw[red!60, thin] (axis cs:-0.40, {sqrt(3*(-0.40)+2.25)}) -- (axis cs:-0.40, 2.0);
\draw[red!60, thin] (axis cs:-0.36, {sqrt(3*(-0.36)+2.25)}) -- (axis cs:-0.36, 2.0);
\draw[red!60, thin] (axis cs:-0.32, {sqrt(3*(-0.32)+2.25)}) -- (axis cs:-0.32, 2.0);
\draw[red!60, thin] (axis cs:-0.28, {sqrt(3*(-0.28)+2.25)}) -- (axis cs:-0.28, 2.0);
\draw[red!60, thin] (axis cs:-0.24, {sqrt(3*(-0.24)+2.25)}) -- (axis cs:-0.24, 2.0);
\draw[red!60, thin] (axis cs:-0.20, {sqrt(3*(-0.20)+2.25)}) -- (axis cs:-0.20, 2.0);
\draw[red!60, thin] (axis cs:-0.16, {sqrt(3*(-0.16)+2.25)}) -- (axis cs:-0.16, 2.0);
\draw[red!60, thin] (axis cs:-0.12, {sqrt(3*(-0.12)+2.25)}) -- (axis cs:-0.12, 2.0);
\draw[red!60, thin] (axis cs:-0.08, {sqrt(3*(-0.08)+2.25)}) -- (axis cs:-0.08, 2.0);
\draw[red!60, thin] (axis cs:-0.04, {sqrt(3*(-0.04)+2.25)}) -- (axis cs:-0.04, 2.0);
\draw[red!60, thin] (axis cs:0.00, 1.5) -- (axis cs:0.00, 2.0);
\draw[red!60, thin] (axis cs:0.04, {sqrt(3*(0.04)+2.25)}) -- (axis cs:0.04, 2.0);
\draw[red!60, thin] (axis cs:0.08, {sqrt(3*(0.08)+2.25)}) -- (axis cs:0.08, 2.0);
\draw[red!60, thin] (axis cs:0.12, {sqrt(3*(0.12)+2.25)}) -- (axis cs:0.12, 2.0);
\draw[red!60, thin] (axis cs:0.16, {sqrt(3*(0.16)+2.25)}) -- (axis cs:0.16, 2.0);
\draw[red!60, thin] (axis cs:0.20, {sqrt(3*(0.20)+2.25)}) -- (axis cs:0.20, 2.0);
\draw[red!60, thin] (axis cs:0.24, {sqrt(3*(0.24)+2.25)}) -- (axis cs:0.24, 2.0);
\draw[red!60, thin] (axis cs:0.28, {sqrt(3*(0.28)+2.25)}) -- (axis cs:0.28, 2.0);
\draw[red!60, thin] (axis cs:0.32, {sqrt(3*(0.32)+2.25)}) -- (axis cs:0.32, 2.0);
\draw[red!60, thin] (axis cs:0.36, {sqrt(3*(0.36)+2.25)}) -- (axis cs:0.36, 2.0);
\draw[red!60, thin] (axis cs:0.40, {sqrt(3*(0.40)+2.25)}) -- (axis cs:0.40, 2.0);
\draw[red!60, thin] (axis cs:0.44, {sqrt(3*(0.44)+2.25)}) -- (axis cs:0.44, 2.0);
\draw[red!60, thin] (axis cs:0.48, {sqrt(3*(0.48)+2.25)}) -- (axis cs:0.48, 2.0);
\draw[red!60, thin] (axis cs:0.52, {sqrt(3*(0.52)+2.25)}) -- (axis cs:0.52, 2.0);
\draw[red!60, thin] (axis cs:0.56, {sqrt(3*(0.56)+2.25)}) -- (axis cs:0.56, 2.0);
\draw[red!60, thin] (axis cs:0.60, {sqrt(3*(0.60)+2.25)}) -- (axis cs:0.60, 2.0);
\draw[red!60, thin] (axis cs:0.64, {sqrt(3*(0.64)+2.25)}) -- (axis cs:0.64, 2.0);
\draw[red!60, thin] (axis cs:0.68, {sqrt(3*(0.68)+2.25)}) -- (axis cs:0.68, 2.0);
\draw[red!60, thin] (axis cs:0.72, {sqrt(3*(0.72)+2.25)}) -- (axis cs:0.72, 2.0);
\draw[red!60, thin] (axis cs:0.76, {sqrt(3*(0.76)+2.25)}) -- (axis cs:0.76, 2.0);
\draw[red!60, thin] (axis cs:0.80, {sqrt(3*(0.80)+2.25)}) -- (axis cs:0.80, 2.0);
\draw[red!60, thin] (axis cs:0.84, {sqrt(3*(0.84)+2.25)}) -- (axis cs:0.84, 2.0);
\draw[red!60, thin] (axis cs:0.88, {sqrt(3*(0.88)+2.25)}) -- (axis cs:0.88, 2.0);
\draw[red!60, thin] (axis cs:0.92, {sqrt(3*(0.92)+2.25)}) -- (axis cs:0.92, 2.0);
\draw[red!60, thin] (axis cs:0.96, {sqrt(3*(0.96)+2.25)}) -- (axis cs:0.96, 2.0);
\draw[red!60, thin] (axis cs:1.00, {sqrt(3*(1.00)+2.25)}) -- (axis cs:1.00, 2.0);

\addplot[blue, thick, domain=-0.25:1]
    {1*sqrt(x+0.25)};

\addplot[orange, thick, domain=-0.5:1]
    {1*sqrt(2*x+1)};

\addplot[green!60!black, thick, domain=-0.75:1]
    {1*sqrt(3*x+2.25)};

\addplot[dashed, gray, domain=0:2] (0.45, x);
\addplot[mark=*, mark size=2pt, fill=blue!100!black, only marks] coordinates {(0.45, 0.83666)};
\addplot[mark=*, mark size=2pt, fill=orange!100!black, only marks] coordinates {(0.45, 1.37840)};
\addplot[mark=*, mark size=2pt, fill=green!80!black, only marks] coordinates {(0.45, 1.89737)};
\addplot[mark=*, mark size=2pt, fill=gray, only marks] coordinates {(0.45, 0)};

\draw[black, thin, ->] (axis cs:0.6, 0.3) -- (axis cs:0.47, 0.06); 
\node[font=\scriptsize] at (axis cs:0.68, 0.37) {$\hat Q_{g_f}''$};

\addplot[dashed, gray, domain=0:2] (0, x);  
\addplot[mark=*, mark size=2pt, fill=blue!100!black, only marks] coordinates {(0, 0.5)};  
\addplot[mark=*, mark size=2pt, fill=orange!100!black, only marks] coordinates {(0, 1.0)};  
\addplot[mark=*, mark size=2pt, fill=green!80!black, only marks] coordinates {(0, 1.5)};  
\addplot[mark=*, mark size=2pt, fill=gray, only marks] coordinates {(0, 0)};

\draw[black, thin, ->] (axis cs:-0.1, 0.35) -- (axis cs:-0.03, 0.05); 
\node[font=\scriptsize] at (axis cs:-0.15, 0.6) {$\hat Q_{g_f}'$};

\draw[blue, thick, ->] (axis cs:0.1, 0.48666) -- (axis cs:0.3, 0.62046);
\draw[orange, thick, ->] (axis cs:0.1, 1.00840) -- (axis cs:0.3, 1.16220);
\draw[green!60!black, thick, ->] (axis cs:0.1, 1.48737) -- (axis cs:0.3, 1.60117);

\node[circle, fill=blue!100!black, inner sep=2pt] at (axis cs:-0.95, 1.65) {};
\node[font=\scriptsize, anchor=west] at (axis cs:-0.90, 1.65) {$\mathrm{SCR} = 1$};

\node[circle, fill=orange!100!black, inner sep=2pt] at (axis cs:-0.95, 1.4) {};
\node[font=\scriptsize, anchor=west] at (axis cs:-0.90, 1.4) {$\mathrm{SCR} = 2$};

\node[circle, fill=green!80!black, inner sep=2pt] at (axis cs:-0.95, 1.1) {};
\node[font=\scriptsize, anchor=west] at (axis cs:-0.90, 1.1) {$\mathrm{SCR} = 3$};

\end{axis}
\end{tikzpicture}
\vspace{-0.8cm}
\caption{Static voltage stability boundaries under different SCR levels. The red-shaded region represents the infeasible operating region from the perspective of static voltage stability for an SCR of 3.}
\label{P_Q_SCR_relationship}
\vspace{-0.5cm}
\end{figure}
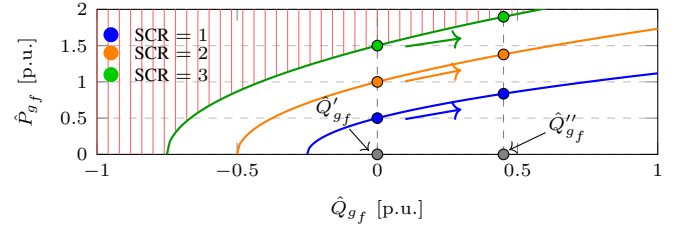

The formulation in \eqref{eq:define_of_original_VS_constraints} takes an SOC form with respect to $\hat{P}_{g_f}$, $\hat{Q}_{g_f}$, and $\hat{Q}_{g_f}+\Gamma_{g_f}$. However, these terms remain highly nonlinear because the impedance ratios, i.e., $|Z_{\Phi({g_f})\Phi({g_f}^{\prime})}|/|Z_{\Phi({g_f})\Phi({g_f})}|$ and $1/|Z_{\Phi({g_f})\Phi({g_f})}|$, involve inverse operations from the grid admittance matrix that depends on the operating points of multiple (V)SGs. Therefore, the offline training in \cite{chu2022voltage} is adopted to linearly approximate these quantities as $z_{g_f}^{{g_f}^\prime} \approx |Z_{\Phi({g_f})\Phi({g_f}^\prime)}|/|Z_{\Phi({g_f})\Phi({g_f})}|$ and $z_{g_f}^1 \approx 1/|Z_{\Phi({g_f})\Phi({g_f})}|$. Details of the training procedure are provided in Appendix~\ref{Approximation of Nonlinear Impedance Ratios}.
  
After the approximation, the terms \eqref{eq:define_VS_P}, \eqref{eq:define_VS_Q}, \eqref{eq:define_Gamma} can now be presented as:
\begin{subequations} \label{eq:P_Q_Tao_approx}
\begin{align}
& \hat{P}_{g_f} = P_{g_f} + \hspace{-0.5cm}
\sum_{\substack{g_f' \in \mathcal{G}_f, g_f' \neq {g_f}}} \hspace{-0.5cm} z_{{g_f}}^{g_f'} P_{g_f'} \label{eq:5a} \\
& \hat{Q}_{g_f} = Q_{g_f} + \hspace{-0.5cm}
\sum_{\substack{g_f' \in \mathcal{G}_f, g_f' \neq {g_f}}} \hspace{-0.5cm} z_{{g_f}}^{g_f'} Q_{g_f'} \label{eq:5b} \\
& \Gamma_{g_f} = \frac{1}{2}z_{{g_f}}^{1} \label{eq:5c}
\end{align}  
\end{subequations}
where $z_{{g_f}}^{g_f'} P_{g_f'}$ and $z_{{g_f}}^{g_f'} Q_{g_f'}$ involve products of binary and continuous variables (i.e., $u_{g_c} P_{g_f'}$, $u_{g_c} Q_{g_f'}$, $\eta_m P_{g_f'}$ and $\eta_m Q_{g_f'}$), which can be linearized using McCormick envelopes \cite{chu2022voltage}.

According to the constraint in \eqref{eq:define_VS_full}, the voltage stability limits at GFL-IBR buses under different SCR levels are illustrated in Fig.~\ref{P_Q_SCR_relationship}. When only the GFL-IBR setpoint is adjusted, such that the SCR remains constant, active power injection can be increased while satisfying the voltage stability constraint through an increase in reactive power injection. For each of the three curves in Fig.~\ref{P_Q_SCR_relationship}, this could be achieved by shifting from unity power factor at $\hat Q_{g_f}'$ to capacitive operation at $\hat Q_{g_f}''$. Alternatively, the operating condition of the main grid can be adjusted to increase SCR, for example by committing more (V)SGs, thereby improving the equivalent impedance seen from the GFL-IBR bus. This would represent moving from the blue curve in Fig.~\ref{P_Q_SCR_relationship} to the yellow or green ones.

\vspace{-0.3cm}
\subsection{Shadow Prices of Static Voltage Stability Services}\label{Shadow Prices for Voltage Stability Constraint}
Based on the expressions \eqref{eq:define_VS_full} and \eqref{eq:P_Q_Tao_approx}, the voltage stability constraint should first be converted into a standard SOC form before defining its associated dual variables \cite{taylor2015convex}: 
\begin{equation}\label{eq:SOC_conver}
    \left\| 
    \begin{bmatrix}
        1 & 0 \\
        0 & 1
    \end{bmatrix}
    \begin{bmatrix}
        \hat{P}_{g_f} \\
        \hat{Q}_{g_f}
    \end{bmatrix}
    \right\|
    \leq 
    [1\;\;1]
    \begin{bmatrix}
        \hat{Q}_{g_f} \\
        \Gamma_{g_f}
    \end{bmatrix}
\end{equation}
where the dual variables are defined as `$\lambda_{1,{g_f}}$' and `$\lambda_{2,{g_f}}$' for the first and second rows in the left-hand side matrix of \eqref{eq:SOC_conver} and `$\mu_{g_f}$' for the vector on the right-hand side. In the dual problem, the following constraint must be set: 
\begin{equation}\label{eq:SOC_shadow_price}
\left\|
\begin{aligned}
 \begin{bmatrix}
\lambda_{1,{g_f}} \\
\lambda_{2,{g_f}}
\end{bmatrix}
\end{aligned}
\right\| \leq \mu_{g_f}
\end{equation}

Consequently, the Lagrangian function of any optimization problem including the voltage stability constraint \eqref{eq:define_of_original_VS_constraints} will include the following term:
\begin{equation} \label{eq:SOC_lagrangian}
    \lambda_{1,{g_f}}\hat{P}_{g_f} + \lambda_{2,{g_f}}\hat{Q}_{g_f} - \mu_{g_f} (\hat{Q}_{g_f} + \Gamma_{g_f})
\end{equation}

Finally, the gradient of \eqref{eq:SOC_lagrangian}, i.e., the KKT stationarity condition that determines the shadow price of the static voltage stability service for GFL-IBR buses, is given by:
\begin{itemize}
    \item Price of $\hat{Q}_{g_f}$: \quad  $-\lambda_{2,{g_f}}+\mu_{g_f}$
    \item Price of $\mathrm{SCR}_{\Phi(g_f)}$ ($=2\Gamma_{g_f}$) enhancement: \quad  $\mu_{g_f}$
\end{itemize}

Note that $\hat{P}_{g_f}$ is not priced as a service for voltage stability, since it is intended to satisfy energy demand and acts as the quantity that consumes voltage stability margin and may deteriorate voltage security, as demonstrated by the simplified form of \eqref{eq:define_VS_full} (i.e., $\hat{P}_{g_f}^{2}  \leq 2\hat{Q}_{g_f}\Gamma_{g_f} + \Gamma_{g_f}^{2}$) and the discussion at the end of Section \ref{Representation of Static Voltage Stability Constraint}. Accordingly, pricing $\lambda_{1,g_f}$ would wrongly remunerate GFL-IBR for keeping $\hat{P}_{g_f}$ within the admissible voltage stability region, which has to be respected regardless. In other words, GFL-IBR should not receive financial compensation merely for avoiding violations of voltage stability constraints.

After establishing the mechanism for assigning shadow prices to the ancillary service, potential schemes for computing these prices are introduced next, together with revenue allocation mechanisms for generators in the market.

\vspace{-0.2cm}
\section{Pricing Methodologies for Handling Non-Convexities in Unit Commitment}\label{Pricing Voltage Stability Services}
This section first formulates a UC model incorporating the static voltage stability constraint, yielding a Mixed-Integer Second-Order Cone Program (MISOCP). It then introduces pricing approaches to determine the service prices while addressing the non-convexity generated from binary variables.

 \vspace{-0.3cm}
\subsection{Static Voltage Stability Constrained UC Problem}\label{Voltage Stability Constrained UC Problem}
A 24-hour UC model with one-hour time intervals is developed to minimize the operating cost of SGs in the day-ahead market, as presented below.
\begin{subequations}\label{eq:MISOCP}
\begin{alignat}{2}
& \min \;& & \sum_{t} \sum_{g_c} \Big( 
\mathrm{c}_{g_c}^\mathrm{nl} u_{g_c,t} 
+ \mathrm{c}_{g_c}^\mathrm{m} P_{g_c,t} 
+ C_{g_c,t}^\mathrm{st} 
+ C_{g_c,t}^\mathrm{sh} \Big) 
\label{eq:VS-cons_UC_obj} \\
& \mathrm{s.t.} & & \mathrm{Voltage~stability~constraint: }\eqref{eq:define_VS_full},~ \forall g_f, t \label{eq:cons_VS} \\
& & & C_{g_c,t}^\mathrm{st} \ge 0,~ \forall g_c, t  \label{eq:cons_st_cost_positive} \\
& & & C_{g_c,t}^\mathrm{sh} \ge 0,~ \forall g_c, t  \label{eq:cons_sh_cost_positive} \\
& & & C_{g_c,t}^\mathrm{st} \ge (u_{g_c,t}-u_{g_c,(t-1)})\mathrm{c}^\mathrm{st}_{g_c},~ \forall g_c,  t  \label{eq:cons_st_cost_lb} \\
& & & C_{g_c,t}^\mathrm{sh} \ge (u_{g_c,(t-1)}-u_{g_c,t})\mathrm{c}^\mathrm{sh}_{g_c},~ \forall g_c, t  \label{eq:cons_sh_cost_lb} \\
& & & u_{g_c,t} \in \{0,1\},~ \forall g_c,  t  \label{eq:binary_SGs} \\
& & & u_{g_c,t} \mathrm{P}_{g_c}^\mathrm{min}  \leq P_{g_c,t} \leq u_{g_c,t}  \mathrm{P}_{g_c}^\mathrm{max},~ \forall g_c,  t  \label{eq:cons_SG_output_active} \\ 
& & & u_{g_c,t} \mathrm{Q}_{g_c}^\mathrm{min}  \leq Q_{g_c,t} \leq u_{g_c,t}  \mathrm{Q}_{g_c}^\mathrm{max},~ \forall g_c,  t  \label{eq:cons_SG_output_reactive} \\ 
& & & 0 \leq P_{g_v,t} \leq \alpha_{g_v,t}\mathrm{P}^{\mathrm{max}}_{g_v},~ \forall g_v, t  \label{eq:cons_VSG_active} \\
& & & \mathrm{Q}_{g_v}^\mathrm{min} \leq Q_{g_v,t} \leq \mathrm{Q}_{g_v}^\mathrm{max},~ \forall g_v, t  \label{eq:cons_VSG_reactive} \\
& & & 0 \leq P_{g_f,t} \leq \alpha_{g_f,t}\mathrm{P}^{\mathrm{max}}_{g_f},~ \forall g_f, t  \label{eq:cons_IBR_active} \\ 
& & & \mathrm{Q}_{g_f}^\mathrm{min} \leq Q_{g_f,t} \leq \mathrm{Q}_{g_f}^\mathrm{max},~ \forall g_f, t  \label{eq:cons_IBR_reactive} \\
& & & P_{g_c,t}^{2}+Q_{g_c,t}^{2} \leq ( {\mathrm{S}_{g_c}^\mathrm{max}})^2,~ \forall g_c,  t  \label{eq:cons_SG_capacity} \\
& & & P_{g_v,t}^{2}+Q_{g_v,t}^{2} \leq ( {\mathrm{S}_{g_v}^\mathrm{max}})^2,~ \forall g_v,  t  \label{eq:cons_VSG_capacity} \\
& & & P_{g_f,t}^{2}+Q_{g_f,t}^{2} \leq ( {\mathrm{S}_{g_f}^\mathrm{max}})^2,~ \forall g_f,  t  \label{eq:cons_IBR_capacity} \\
& & & \sum_{i \in \{\Phi(g_c), \Phi(g_v), \Phi(g_f)\}} \hspace{-1cm} P_{i,t} = \sum_i \mathrm{P}_{i,t}^\mathrm{D},~ \forall t \label{eq:cons_balance_active} \\
& & & \sum_{i \in \{\Phi(g_c), \Phi(g_v), \Phi(g_f)\}} \hspace{-1cm} Q_{i,t} = \sum_i \mathrm{Q}_{i,t}^\mathrm{D},~\forall t \label{eq:cons_balance_reactive}
\end{alignat}
\end{subequations} 
where the objective \eqref{eq:VS-cons_UC_obj} includes no-load, generation and startup/shutdown costs of SGs. Eq.~\eqref{eq:cons_VS} regulates relevant variables to maintain the voltage stability of GFL-IBR buses; Eqs.~\eqref{eq:cons_st_cost_positive}-\eqref{eq:cons_sh_cost_lb} define startup/shutdown costs; Eq.~\eqref{eq:binary_SGs} imposes the binary nature of commitments; Eqs.~\eqref{eq:cons_SG_output_active}-\eqref{eq:cons_SG_output_reactive}, \eqref{eq:cons_VSG_active}-\eqref{eq:cons_VSG_reactive}, \eqref{eq:cons_IBR_active}-\eqref{eq:cons_IBR_reactive} limit the active/reactive power output of SGs, VSGs and GFL-IBR, respectively; Eqs.~\eqref{eq:cons_SG_capacity}-\eqref{eq:cons_IBR_capacity} require the active/reactive power injections from generators to be within the apparent power capacity; Eqs.~\eqref{eq:cons_balance_active}-\eqref{eq:cons_balance_reactive} balance the active/reactive power of the system.

\subsection{Methodologies for Pricing Static Voltage Stability Services}
Before presenting the specific pricing models, the relevant economic principles underlying the proposed market framework are introduced as follows \cite{gribik2007market,ruiz2012pricing}:
\begin{enumerate}
    \item Within this market framework, competitive producers submit offers to the Independent System Operator (ISO), which determines market prices and production quantities to achieve market clearing.
    \item Competitive producers are price takers that accept market clearing prices for revenue and profit settlement and cannot readjust their submitted offers afterward.
    \item The prices provide revenue adequacy for service‑providing units. Although these prices may not guarantee a competitive partial equilibrium and some generating units may still experience opportunity costs, the dispatched producers achieving non negative profits would operate in a quasi partial equilibrium, in the sense that none of them would have an incentive to exit the market, though individual units may still prefer alternative output levels under the posted prices to maximize profits.
    \item Discrete decision variables can produce circumstances where there would be no exact prices that support the optimal system dispatch. 
\end{enumerate}

The main challenge in pricing voltage stability services arises from the non-convexity introduced by the commitment decisions of conventional SGs. To address this challenge, this paper examines three alternative pricing methods, namely `marginal unit', `restricted', and `dispatchable'.

\subsubsection{Marginal Unit Pricing Method}\label{Marginal Unit Pricing Method}
This pricing process does not rely on duality theory and can therefore handle cases in which the gradient of the Lagrangian function is undefined due to discrete non-convexities. It quantifies the ancillary service price for each generator based on variations in the system operating cost \eqref{eq:VS-cons_UC_obj} resulting from that generator’s contribution to the voltage stability constraint. The detailed implementation steps of modifying UC model for evaluating cost variations are presented in Appendix~\ref{Procedure of Marginal Unit Pricing}.

Let `$f^*$' and `$f_{ \{g_c,g_v,g_f\} }$' denote the optimal objective values of the original and modified MISOCPs, respectively. The ancillary service prices for SGs, VSGs, and GFL-IBR are obtained through the following comparison:
\begin{equation}\label{eq:marginal_unit_pricing_objs}
    f_{ \{g_c,g_v,g_f\} } - f^*
\end{equation}
where \eqref{eq:marginal_unit_pricing_objs} represents the incremental cost incurred by dispatching alternative resources to compensate for the lost contribution to voltage stability.

Although this method can handle non-convexities and is conceptually intuitive, its application to large-scale systems would require repeated solution of the optimization model. Furthermore, it cannot capture the interactions among different types of services in the market.

\subsubsection{Restricted Pricing Method}\label{Restricted Pricing Method}
To achieve coordinated market clearing while capturing interactions among different services and computing shadow prices, the restricted pricing approach based on duality theory solves the MISOCP twice. First, the original model is solved, yielding the UC schedule, which is recorded as the optimal value `$u_{g_c}^*$'. Then, the model is solved again with the binary variables relaxed and confined to take the previously obtained value:
\begin{equation}\label{eq:restricted_define}
    u_{g_c}=u_{g_c}^*, \quad \eta_m=\eta_m^*
\end{equation}
Thus, a new term is added to the Lagrangian function \eqref{eq:SOC_lagrangian}:
\begin{equation}
    \lambda_{g_c,\mathrm{commit}}(u_{g_c}^*-u_{g_c})
\end{equation}
where `$\lambda_{g_c,\mathrm{commit}}$' denotes the commitment price, representing a side payment for units associated with their commitment decisions. It also serves as a part of uplift payment for relevant generators to cover profit losses \cite{gribik2007market}. 

However, generating units without commitment variables will not capture commitment prices, such as IBR. This limitation was also noted in the pricing of other ancillary services, such as inertia \cite{badesa2022assigning}. In addition, clear rules are required to clarify who shall bear such uplift payments in the market, as different uplift‑cost‑allocation mechanisms yield distinct trade‑offs between market fairness and efficiency \cite{hytowitz2020impacts}. Socializing uplift costs across all system loads spreads the financial burden broadly yet may weaken individual incentives. Alternatively, assigning charges to specific market participants based on their scheduling‑related deviations or contributions can better align cost responsibility with the factors leading to uplift, yet would introduce concerns regarding fairness and implementation complexity.

\subsubsection{Dispatchable Pricing Method}\label{Dispatchable Pricing Method}
To attempt to avoid make-whole payments and settle market participants' revenues purely through market clearing prices, the dispatchable pricing method calculates shadow prices by relaxing all binary variables and assuming that all units are fully dispatchable. The integer requirement $u_{g_c}\in\{0, 1\}$ is therefore replaced by:
\begin{equation}
    0 \leq u_{g_c} \leq 1
\end{equation}
The nonlinear terms such as $u_{g_c} P_{{g_f}'}$ mentioned in \eqref{eq:P_Q_Tao_approx} then become products of continuous variables, which are non-convex. Although McCormick envelopes could be used to construct the convex hull for these bilinear terms, it is typically coarse and would overly relax the model. 

In summary, the two duality‑based pricing schemes in this work do not apply duality on the original MISOCP. Instead, restricted pricing fixes binary variables at pre‑computed optimal dispatch values, while dispatchable pricing relaxes binary variables into continuous bounds. Both approaches in fact yield convex SOCP where strong duality is valid for shadow‑price calculation, at the cost of respective practical limitations as discussed in Section~\ref{Case Studies}.

\vspace{-0.3cm}
\subsection{Settlement of Market Participants}\label{Settlement of Market Participants}
This section specifies the market settlement rules, clarifying how revenues and payments (particularly for voltage stability services) are determined for market buyers and sellers.

\subsubsection{Revenue and Profit of Generators}
In this system, generators rely on revenues from energy and voltage stability service markets in an attempt to cover their operating costs.
Since (V)SGs and GFL-IBR deliver distinct forms of voltage stability support (as discussed at the end of Section \ref{Representation of Static Voltage Stability Constraint}), their ancillary service revenues are separately formulated as follows:
\begin{subequations} \label{eq:VS_revenue_IBR_VSGs}
\begin{align}
& \underbrace{\mu_{g_f} ~ \frac{1}{\Delta|Z_{\Phi({g_f})\Phi({g_f})}|}}_{\mathrm{Voltage~stability~service~revenue~for~each~(V)SG}}
\label{eq:VS_revenue_(V)SGs} \\
& \underbrace{
- (\lambda_{2,{g_f}} - \mu_{g_f})Q_{g_f} 
- \hspace{-0.6cm} \sum_{\substack{g_f' \in \mathcal{G}_f, g_f' \neq {g_f}}} \hspace{-0.6cm}
(\lambda_{2,g_f'} - \mu_{g_f'})\frac{|Z_{\Phi(g_f')\Phi({g_f})}|}{|Z_{\Phi(g_f')\Phi(g_f')}|}Q_{g_f}
}_{\mathrm{Voltage~stability~service~revenue~for~each~GFL-IBR}}
\label{eq:VS_revenue_IBR_Q}
\end{align}
\end{subequations}
where \eqref{eq:VS_revenue_(V)SGs} is associated with SCR enhancement, which is captured by the impedance variation resulting from the change of their operating state. Eq.~\eqref{eq:VS_revenue_IBR_Q} is associated with reactive power provision, aimed at maintaining voltage stability at local and remote GFL-IBR buses. 

Additionally, since the restricted pricing method introduces commitment prices, SGs also capture the reward:
\begin{equation}\label{eq:SG_capture_commit_price}
    \lambda_{g_c,\mathrm{commit}} \cdot u_{g_c}
\end{equation}

Based on the above revenues for the units, their final profits are expressed as follows:
\begin{subequations} \label{eq:profit_units}
\begin{gather}
\hspace{-0.1cm}\underbrace{ \lambda^\mathrm{E} P_{g_c} \hspace{-0.05cm} + \hspace{-0.05cm} \eqref{eq:VS_revenue_(V)SGs} \hspace{-0.05cm} + \hspace{-0.05cm} \eqref{eq:SG_capture_commit_price} \hspace{-0.05cm} - \hspace{-0.05cm} ( \mathrm{c}_{g_c}^\mathrm{nl} u_{g_c} 
 +  \mathrm{c}_{g_c}^\mathrm{m} P_{g_c} 
 +  C_{g_c}^\mathrm{st} 
 +  C_{g_c}^\mathrm{sh}  ) }_{\mathrm{Profit~for~each~SG}} \label{eq:profit_SG} \\
\underbrace{ \lambda^\mathrm{E} P_{g_v} + \eqref{eq:VS_revenue_(V)SGs} }_{\mathrm{Profit~for~each~VSG}} \label{eq:profit_VSG}  \\
\underbrace{ \lambda^\mathrm{E} P_{g_f} + \eqref{eq:VS_revenue_IBR_Q} }_{\mathrm{Profit~for~each~GFL-IBR}} \label{eq:profit_IBR} 
\end{gather}
\end{subequations}
where `$\lambda^\mathrm{E}$' is the energy price, a dual variable associated with the active power balance constraint \eqref{eq:cons_balance_active}. Eq.~\eqref{eq:profit_SG} incorporates operating costs of SGs. Such costs are excluded for VSGs and GFL-IBR, since the RES they utilize is regarded as cost-free. For the dispatchable method, \eqref{eq:SG_capture_commit_price} is removed from \eqref{eq:profit_SG}.

Note that the revenue and profit calculations herein do not apply to marginal unit pricing, as this approach directly allocates hourly ancillary service revenues to generators based on changes in system operating costs, rather than dual variables.

\subsubsection{Entities Who Pay for Ancillary Services}
Energy market payments are collected from consumers proportional to their electricity consumption, which is not further elaborated here. For voltage stability services, two distinct cost allocation mechanisms are discussed below.

First, the socialized allocation distributes the total ancillary service bill among all consumers according to the ratio of their energy consumption to the total system load, as formulated in:
\begin{equation}
\label{eq:social_allocation}
\mathrm{Bill}^\mathrm{AS,chrg}_{i,t}
= \mathrm{Bill}^\mathrm{AS,tot}_t
\cdot \frac{\mathrm{P}^\mathrm{D}_{i,t}}{\sum_{i\in \mathcal{I}} \mathrm{P}^\mathrm{D}_{i,t}},\quad \forall i, t
\end{equation}
where `$\mathrm{Bill}^\mathrm{AS,tot}_t$' denotes the total ancillary service revenue paid to all service‑providing units. `$\mathrm{Bill}^\mathrm{AS,chrg}_{i,t}$' represents the ancillary service cost charged to the load at node $i$.

Second, the liability-triggered allocation assigns costs based on specific triggering sources. On the one hand, when GFL-IBR need higher levels of SCR, they shall cover the ancillary service payments for supporting services provided by (V)SGs. On the other hand, when the demand side drives GFL-IBR to increase active power injection, the required reactive power coordination for voltage stability may be triggered by load consumption; accordingly, the corresponding ancillary service costs should be allocated to consumers.

These two allocation schemes exhibit distinct advantages and limitations \cite{matamala2024cost}. Socialized cost allocation is conceptually clear and practically implementable. However, the marginal increases or decreases of electricity demand may not trigger the ancillary service, which may lead to overcharging of certain consumers. By contrast, liability‑triggered allocation requires very sophisticated settlement rules. Poorly designed allocation criteria may result in market entities paying more than the actual cost required to satisfy system‑security requirements. This challenge would be aggravated by network impedance coupling in this market, which complicates the attribution of each component of ancillary service provision. 

Developing dedicated cost allocation methodologies is still an active and separate research area, which falls outside the scope of this paper and is left for future work.

\section{Case Studies}\label{Case Studies}
For the adopted power system, three distinct pricing methods for handling non-convexities are examined to assess whether they can yield market prices that support optimal system operation, along with a discussion of their respective limitations. Prior to the pricing analysis, system parameters and configurations are first defined, and the validity of the approximated voltage stability constraints needs to be verified.

Note that in Fig.~\ref{fig:Shadow prices of service_restricted} and Fig.~\ref{fig:Shadow prices of service_dispatchable}, the unit of y-axis refers to the shadow prices of the two voltage stability ancillary services (SCR enhancement and reactive power support). The term `price per service' is only used for compact illustration to facilitate comparison between different ancillary service components, and does not represent a distinct economic pricing paradigm. Corresponding physical units are specified in each figure caption.

\vspace{-0.3cm}
\subsection{Test System Setting}


 \begin{table}[t]
\centering
\caption{Operation Characteristics of Thermal Units}
         \vspace{-0.3cm}
\setlength{\tabcolsep}{4pt}
{\fontsize{8pt}{12pt}\selectfont
\begin{tabular}{lcccccc}
\toprule
Bus & 2 & 3 & 4 & 5 & 27 & 30 \\
\midrule
\(\mathrm{c}_{g_c}^\mathrm{nl}\) (\texteuro/h) & 348.60 & 300.20 & 275.20 & 218.60 & 198.00 & 171.40 \\
\(\mathrm{c}_{g_c}^\mathrm{m}\) (\texteuro/MWh) & 6.20 & 6.91 & 10.47 & 12.28 & 13.53 & 15.36 \\
\(\mathrm{c}_{g_c}^\mathrm{st}\) (\texteuro) & 2000 & 1250 & 925 & 720 & 550 & 310  \\
\(\mathrm{c}_{g_c}^\mathrm{sh}\) (\texteuro) & 500 & 285 & 185 & 144 & 120 & 100 \\
\(\mathrm{P}_{g_c}^\mathrm{min}\) (MW) & 43.87 & 38.40 & 20.13 & 8.87 & 8.67 & 3.87 \\
\(\mathrm{P}_{g_c}^\mathrm{max}\) (MW) & 87.80 & 76.80 & 50.40 & 44.47 & 43.33 & 38.40 \\
\bottomrule
\end{tabular}
}
\label{table:units_para}
\vspace{-0.3cm}
\end{table}

\begin{table}[t]
\centering
\caption{MAPE of Approximation Errors for Nonlinear Terms in Static Voltage Stability Constraints}
         \vspace{-0.3cm}
\setlength{\tabcolsep}{6pt} 
{\fontsize{8pt}{12pt}\selectfont
\begin{tabular}{lcccc}
\toprule
Approximated terms & $\frac{|Z_{23,24}|}{|Z_{23,23}|}$ & $\frac{|Z_{24,23}|}{|Z_{24,24}|}$ & $\frac{1}{|Z_{23,23}|}$ & $\frac{1}{|Z_{24,24}|}$ \\
\midrule
\hspace{0.6cm} MAPE & $0.34$\% & $0.17$\% & $3.16$\% & $2.52$\% \\
\bottomrule
\end{tabular}
}
\label{table:Approximation_Errors}
        \vspace{-0.4cm}
\end{table}

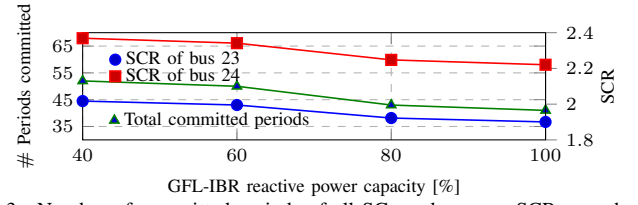
\begin{figure}[t]
\centering
\begin{adjustbox}{minipage=[t]{0.87\linewidth},center}
\begin{tikzpicture}

\begin{axis}[
    width=1\linewidth, height=3cm,
    xlabel={\scriptsize GFL-IBR reactive power capacity [\%]},
    ylabel={\scriptsize $\#$ Periods committed},
    label style={font=\scriptsize}, tick label style={font=\scriptsize},
    xmin=40, xmax=100,
    ymin=30, ymax=70,
    xtick={40,60,80,100},
    ytick={35,45,55,65,75},
    grid=both, grid style={dashed, gray!60},
    legend style={
        font=\scriptsize, 
        fill=none, 
        draw=none, 
        at={(0.28,0)}, anchor=south
    },
    legend cell align={left},
    legend image post style={scale=1}, 
    legend image code/.code={%
        \draw[mark repeat=1,mark phase=1,#1] plot coordinates {(0,0)}; 
    },
    axis y line*=left,
    axis x line*=bottom,
    axis lines=box
]

\addplot+[   color=green!50!black, mark=triangle*, line width=0.6pt]
    coordinates {
        (40,52) (60,50) (80,43) (100,41)
    };
\addlegendentry{Total committed periods}
\end{axis}

\begin{axis}[
    width=\linewidth, height=3cm,
    xmin=40, xmax=100,
    ymin=1.80, ymax=2.40,
    axis y line*=right,
    axis x line=none,
    ylabel={\scriptsize SCR },
    xtick={40,60,80,100},
    ytick={1.80,2.00,2.20,2.40},
    tick label style={font=\scriptsize},
    tick align=outside, 
    ylabel style={at={(axis description cs:1.10,0.5)}},
    legend style={
        font=\scriptsize, 
        fill=none, 
        draw=none, 
        at={(0.37,0.93)}, anchor=north east,
                row sep=-4pt,
    },
    legend cell align={left},
    legend image post style={scale=1}, 
    legend image code/.code={%
        \draw[mark repeat=1,mark phase=1,#1] plot coordinates {(0,0)}; 
    }
]

\addplot+[   color=blue, mark=*, line width=0.6pt]
    coordinates {
        (40,2.017219707615551) (60,1.9952014198490378)
        (80,1.9222473458737994) (100,1.9004555259486977)
    };
\addlegendentry{SCR of bus 23}

\addplot+[   color=red, mark=square*, line width=0.6pt]
    coordinates {
        (40,2.3689216246506204) (60,2.3413134911646214)
        (80,2.247694031506115) (100,2.2204490942062676)
    };
\addlegendentry{SCR of bus 24}

\end{axis}

\end{tikzpicture}
\end{adjustbox}
\vspace{-0.9cm}
\caption{Number of committed periods of all SGs and average SCR over the 24-h operation, under various GFL-IBR reactive power capacity in the system.}
\label{fig:UC_strength}
\vspace{-0.5cm}
\end{figure}

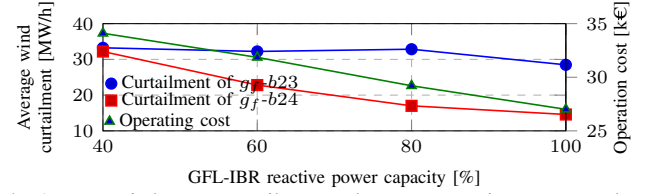
\begin{figure}[t]
\centering
\begin{adjustbox}{minipage=[t]{0.87\linewidth},center}
\begin{tikzpicture}

\begin{axis}[
    width=1\linewidth, height=3cm,
    xlabel={\scriptsize GFL-IBR reactive power capacity [\%]},
     ylabel={\scriptsize \parbox[c]{2.5cm}{\centering Average wind curtailment {[MW/h]}}}, 
    ylabel style={align=center, yshift=-0.3cm},
    label style={font=\scriptsize}, tick label style={font=\scriptsize},
    xmin=40, xmax=100,
    ymin=10, ymax=40,
    xtick={40,60,80,100},
    ytick={10,20,30,40},
    xlabel style={at={(axis description cs:0.5,-0.15)}},
    ylabel style={at={(axis description cs:-0.07,0.5)}},
    grid=both, grid style={dashed, gray!60},
    legend style={
        font=\scriptsize, 
        fill=none, 
        draw=none, 
        at={(0.22,0.1)}, anchor=south,
        row sep=-4pt,
    },
     legend cell align={left},
    legend image post style={scale=1}, 
    legend image code/.code={%
        \draw[mark repeat=1,mark phase=1,#1] plot coordinates {(0,0)}; 
    },
    axis y line*=left,
    axis x line*=bottom,
     axis lines=box
]

\addplot+[   color=blue, mark=*, line width=0.6pt]
    coordinates {
        (40,33.23596550273205) (60,32.218198016660075)
        (80,32.84420804361674) (100,28.4756900592576)
    };
\addlegendentry{Curtailment of $g_f$\textrm{-}$b23$ }

\addplot+[   color=red, mark=square*, line width=0.6pt]
    coordinates {
        (40,32.196835348602484) (60,22.81430559198199)
        (80,17.004117075035005) (100,14.582971003355643)
    };
\addlegendentry{Curtailment of $g_f$\textrm{-}$b24$ }

\end{axis}

\begin{axis}[
    width=1\linewidth, height=3cm,
    xmin=40, xmax=100,
    ymin=25, ymax=35,
    axis y line*=right,
    axis x line=none,
    ylabel={\scriptsize Operation cost [k\texteuro]},
    tick align=outside, 
    ylabel style={align=center, yshift=0.1cm}, 
    ytick={25,30,35},
    xtick={40,60,80,100},
    tick label style={font=\scriptsize},
    ylabel style={at={(axis description cs:1.10,0.5)}},
    legend style={
        font=\scriptsize, 
        fill=none, 
        draw=none, 
        at={(0.305,0.27)}, anchor=north east,
    },
    legend image post style={scale=1}, 
    legend image code/.code={%
        \draw[mark repeat=1,mark phase=1,#1] plot coordinates {(0,0)}; 
    },
]

\addplot+[   color=green!50!black, mark=triangle*, line width=0.6pt]
    coordinates {
        (40,34.10094226529944) (60,31.855587482347484)
        (80,29.2061105674687) (100,27.009171210069584)
    };
\addlegendentry{Operating cost}
\end{axis}

\end{tikzpicture}
\end{adjustbox}
\vspace{-0.9cm}
\caption{Average wind power curtailment and system operation cost over the 24-h operation, under various GFL-IBR reactive power capacity in the system.}
       \vspace{-0.4cm}
\label{fig:wind_curt_and_cost}
\end{figure}

Case studies based on a modified IEEE 30-bus system are conducted to evaluate the proposed schemes for pricing services. 
Wind turbines are installed at buses \{1, 23, 24\}, with the first operating as a VSG and the remaining ones operating as GFL-IBR. Conventional SGs are located at buses \{2, 3, 4, 5, 27, 30\}. The operating parameters of generators are listed in Table~\ref{table:units_para}. The total wind capacity is 270 MW, while system load varies within the range $\in [176, 451]$ MW. The base power is $\mathrm{S}_{\mathrm{B}}=100~\mathrm{MVA}$. A notation is set to denote the SG at bus 2 as $g_c$-$b2$. The code implementation in \texttt{Julia-JuMP} used in this work is publicly available in \cite{Code}. Numerical experiments were performed on an Apple MacBook Air (M1, 2020) with \texttt{Gurobi~12.0.1} as the MISOCP solver.

\vspace{-0.3cm}
\subsection{Validation of Static Voltage Stability Constraints}\label{Validation of Voltage Stability Constraints}
This subsection validates the approximated voltage stability constraint. It first examines the errors produced from the approximation process. Afterwards, a sensitivity analysis of the system operation with respect to reactive power injection of GFL-IBR is performed.

\subsubsection{Approximation Errors of Static Voltage Stability Constraints}
For the test system, the form $z_{g_f}^{{g_f}',\mathrm{II}}$ (resp. $z_{g_f}^{1,\mathrm{I}}$) in \eqref{eq:approx_z_delta} is adopted to approximate the actual $\mathbf{z}_{23}^{24}$ and $\mathbf{z}_{24}^{23}$ (resp. $\mathbf{z}_{23}^{1}$ and $\mathbf{z}_{24}^{1}$), as only this choice satisfies the acceptable approximation accuracy criterion evaluated using the Mean Absolute Percentage Error (MAPE). As reported in Table~\ref{table:Approximation_Errors}, the overall approximation MAPE remains below the acceptable threshold of $5\%$ defined in \cite{chu2022voltage}, indicating a feasible training result. 

\subsubsection{System Sensitivity to GFL-IBR Reactive Capacity}
Variations in system operation are investigated by progressively increasing the overall reactive power contribution from GFL-IBR in increments of $20\%$, from $40\%$ to $100\%$. As shown in Fig.~\ref{fig:UC_strength}, increased reactive support alleviates the need for SGs to remain online to enhance SCR for securing the voltage stability. This facilitates higher wind power delivery and curtails thermal generation costs, thereby lowering total system operating costs, as seen in Fig.~\ref{fig:wind_curt_and_cost}. When the IBR are allowed to utilize $40\%$–$80\%$ of their reactive power capacity, the injected wind power of $g_f\textrm{-}b23$ remains constant. This is because bus 23 has a lower SCR than bus 24, which constrains the maximum active power output at this bus, i.e., $P_{g_f\textrm{-}b23}$. Nevertheless, its equivalent output, $\hat{P}_{g_f\textrm{-}b23}$ as defined in \eqref{eq:5a}, still exhibits an increasing trend.

With the validated voltage stability constraint, the proposed pricing schemes for handling non-convexities are examined assuming full reactive power utilization by GFL-IBR.

\vspace{-0.5cm}
\subsection{Pricing Ancillary Services by Marginal Unit Method}
The ancillary service prices for each generator are sequentially determined by comparing the system cost of the original MISOCP with that of a modified MISOCP, where the voltage stability contribution of the corresponding generator is removed, as the procedure in Section~\ref{Marginal Unit Pricing Method}.

\begin{figure}[t]
\centering
\begin{adjustbox}{minipage=[t]{1\linewidth},center}
\begin{tikzpicture}
\begin{axis}[
    width=1\linewidth, height=3cm,
    xlabel={\scriptsize Time [h]},
     ylabel={\scriptsize \parbox[c]{2cm}{\centering Ancillary service prices {[k\texteuro/unit]}}}, 
    ylabel style={align=center, yshift=-0.3cm},
    label style={font=\scriptsize}, 
    tick label style={font=\scriptsize},
    xmin=1, xmax=24,
    ymin=0, ymax=1.5,
    ytick distance=0.3,
    xtick={4,8,12,16,20},
    grid=both, grid style={dashed, gray!60},
    legend style={
        font=\scriptsize, 
        fill=none, 
        draw=none, 
        at={(0.42,0.45)}, anchor=south west,
        legend columns=2,
        row sep=-2pt
    },
    legend cell align={left},
    legend image post style={scale=1},
    legend image code/.code={
        \draw[mark repeat=1,mark phase=1,#1] plot coordinates {(0,0)};
    },
    axis lines=box
]

\addplot+[color=blue, solid, mark=o, mark size=1.5pt, line width=0.6pt]
coordinates {
(1,0)
(2,0)
(3,0)
(4,0)
(5,0)
(6,0)
(7,0)
(8,0)
(9,0)
(10,0.052954517037837604)
(11,0.05295451555829777)
(12,0.05295451566555357)
(13,0.05295451723706242)
(14,0.05295457883783433)
(15,0.05295451577371)
(16,0.05295451468594388)
(17,0)
(18,0.9428565551597253)
(19,0.26430532627106027)
(20,0)
(21,0)
(22,0)
(23,0)
(24,0)
};
\addlegendentry{$g_c$-$b2$} 

\addplot+[color=red, solid, mark=o, mark size=1.5pt, line width=0.6pt]
coordinates {
(1,0)
(2,0)
(3,0)
(4,0)
(5,0)
(6,0)
(7,0)
(8,0)
(9,0)
(10,0)
(11,0)
(12,0)
(13,0)
(14,0)
(15,0)
(16,0)
(17,0)
(18,0.30306095692225153)
(19,0.08219711477682245)
(20,0)
(21,0)
(22,0)
(23,0)
(24,0)
};
\addlegendentry{$g_c$-$b3$}

\addplot+[ color=green!50!black, solid, mark=o, mark size=1.5pt, line width=0.6pt]
coordinates {
(1,0)
(2,0)
(3,0)
(4,0)
(5,0)
(6,0)
(7,0.05634)
(8,0.010132)
(9,0.099917)
(10,1.38835)
(11,0.129481)
(12,0.160121)
(13,0.15367)
(14,0.148832)
(15,0.05772)
(16,0.156089)
(17,0)
(18,0.143446)
(19,0.008538)
(20,0)
(21,0.503264)
(22,0.018155)
(23,0)
(24,0)
};
\addlegendentry{$g_c$-$b27$}

\addplot+[color=orange, solid, mark=o, mark size=1.5pt, line width=0.6pt]
coordinates {
(1,0.016061)
(2,0)
(3,0)
(4,0.05178)
(5,0)
(6,0)
(7,1.047483)
(8,0)
(9,0.260712)
(10,0.25572)
(11,0.254208)
(12,0.259433)
(13,0.258333)
(14,0.257508)
(15,0.24197)
(16,0.258745)
(17,0)
(18,0.689501)
(19,0.389669)
(20,0.077458)
(21,0.260728)
(22,0.040635)
(23,0.144)
(24,0)
};
\addlegendentry{$g_v$-$b1$}

\end{axis}
\end{tikzpicture}
\end{adjustbox}
\vspace{-0.9cm}
\caption{Voltage stability service prices received by different (V)SGs under the marginal unit pricing method. Prices for the remaining SGs are zero.}
\label{fig:VSGs_op_state_price}
      \vspace{-0.2cm}
\end{figure}
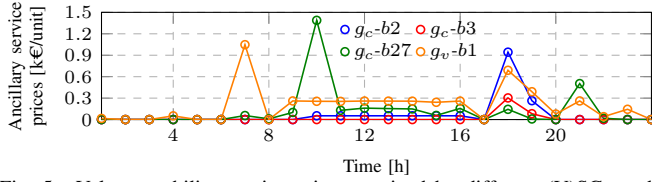

\begin{table}[t]
\caption{Operating Costs and Ancillary Service Remuneration to Generators Using Marginal Unit Pricing Method (k\texteuro/day)}
\vspace{-0.3cm}
\centering
\setlength{\tabcolsep}{3.5pt}
{\fontsize{8pt}{12pt}\selectfont
\begin{tabular}{cccc}
\toprule
\multirow{2}{*}{Units} & \multirow{2}{*}{Costs} & \multicolumn{2}{c}{Remuneration by voltage stability services} \\
\cmidrule(lr){3-4}
& & \hspace{0.5cm} $\hat{Q}_{g_f}$ & \hspace{0.5cm} $\mathrm{SCR}_{\Phi(g_f)}$  \\
\midrule
$g_v$-$b1$    & $\mathrm{0}$ & \hspace{0.5cm} $\mathrm{0}$    & $\mathrm{4.76}$  \\ \hline
$g_c$-$b2$    & $\mathrm{14.39}$ & \hspace{0.5cm} $\mathrm{0}$    & $\mathrm{1.58}$  \\
$g_c$-$b3$    & $\mathrm{3.64}$ & \hspace{0.5cm} $\mathrm{0}$    & $\mathrm{0.39}$  \\
$g_c$-$b27$   & $\mathrm{8.98}$ & \hspace{0.5cm} $\mathrm{0}$    & $\mathrm{3.03}$  \\
other SGs     & $\mathrm{0}$    & \hspace{0.5cm} $\mathrm{0}$    & $\mathrm{0}$     \\
\midrule
$g_f$-$b23$   & $\mathrm{0}$ & \hspace{0.5cm} $\mathrm{0.079}$ & $\mathrm{0}$    \\
$g_f$-$b24$   & $\mathrm{0}$ & \hspace{0.5cm} $\mathrm{1.37}$ & $\mathrm{0}$    \\
\bottomrule
\end{tabular}
}
\label{tab:unit_revenue_marginal_unit}
\vspace{-0.3cm}
\end{table}

\begin{figure}[!t]
\centering
\begin{adjustbox}{minipage=[t]{1\linewidth},center}
\begin{tikzpicture}
\begin{axis}[
    width=0.975\linewidth, height=3cm,
    xlabel={\scriptsize Time [h]},
     ylabel={\scriptsize \parbox[c]{2cm}{\centering Ancillary service prices {[k\texteuro/unit]}}}, 
    ylabel style={align=center, yshift=-0.3cm},
    label style={font=\scriptsize},
    tick label style={font=\scriptsize},
    xmin=1, xmax=24,
    ymin=0, ymax=1.2,
    ytick distance=0.4,
    xtick={4,8,12,16,20},
    grid=both, 
    grid style={dashed, gray!60},
    legend style={
        font=\scriptsize, 
        fill=none, 
        draw=none, 
        at={(0.16,0.36)}, anchor=south,
        legend columns=1,
        row sep=-1.5pt
    },
    legend cell align={left},
    legend image post style={scale=1},
    legend image code/.code={%
        \draw[mark repeat=1,mark phase=1,#1] plot coordinates {(0,0)};
    },
    axis lines=box
]

\addplot+[color=red, solid, mark=o, mark size=1.5pt, line width=0.6pt]
coordinates {
(1,0)
(2,0)
(3,0)
(4,0)
(5,0)
(6,0)
(7,0)
(8,0)
(9,0)
(10,0.053647195475961325)
(11,0)
(12,0)
(13,0)
(14,0)
(15,0)
(16,0)
(17,0)
(18,0)
(19,0)
(20,0.02556562812130187)
(21,0)
(22,0)
(23,0)
(24,0)
};
\addlegendentry{$Q_{g_f\textrm{-}b23}$}

\addplot+[color=green!50!black, solid, mark=star, mark size=1.5pt, line width=0.6pt]
coordinates {
(1,0)
(2,0)
(3,0)
(4,0)
(5,0)
(6,0)
(7,0)
(8,0)
(9,0)
(10,0.053647)
(11,0)
(12,0)
(13,0)
(14,0)
(15,1.033662)
(16,0)
(17,0)
(18,0.286866)
(19,0)
(20,0)
(21,0)
(22,0)
(23,0)
(24,0)
};
\addlegendentry{$Q_{g_f\textrm{-}b24}$}

\end{axis}
\end{tikzpicture}
\end{adjustbox}
\vspace{-0.9cm}
\caption{Voltage stability service prices received by GFL-IBR under the marginal unit pricing method.}
\label{fig:IBR_P_Q_price_marginal_unit}
       \vspace{-0.5cm}
\end{figure}
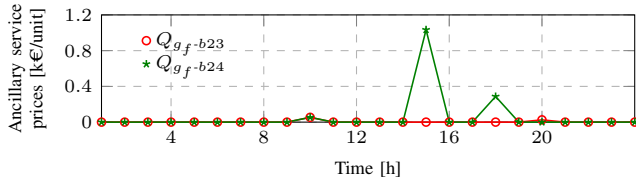

The hourly ancillary service prices for (V)SGs are depicted in Fig.~\ref{fig:VSGs_op_state_price}, reflecting the economic value of the SCR enhanced by these units. The remuneration received by $g_c$–$b2$, $g_c$–$b27$ and $g_v$–$b1$ is noticeably higher than that of $g_c$–$b3$ (as also summarized in Table~\ref{tab:unit_revenue_marginal_unit}, which aggregates hourly remuneration over the day), primarily because the voltage stability-related terms coupled with these units contribute more to system voltage stability. When these terms are eliminated to evaluate ancillary service prices, the system operator has to dispatch more additional resources to compensate for the lost contribution, thereby increasing the system operating cost. Except for these operating generators, the remaining SGs receive no ancillary service payments. This may be attributed either to their not being selected for operation or to their voltage stability contributions being too minor to influence UC outcomes.

Fig.~\ref{fig:IBR_P_Q_price_marginal_unit} illustrates the economic value of reactive power from GFL-IBR in supporting voltage stability. As shown, the reactive power purely enhances voltage stability, leading to non-negative service prices throughout the market horizon. At 15:00 and 18:00, the ancillary service revenue of $g_f\textrm{-}b24$ surges from zero rather than gradually reaching the peak. This occurs because marginal unit pricing completely strips continuous reactive power contributions from voltage stability constraints, in the sense that reactive power is treated as a discrete service, preventing its exact marginal contribution to the system from being captured.

As expected, a key limitation of the marginal unit pricing method is that it cannot be integrated with other services into a unified pricing framework based on duality theory. As a result, the energy price cannot be determined. This limitation is particularly important in the present case, where ancillary service revenues alone are insufficient to recover generator operating costs, as shown in Table~\ref{tab:unit_revenue_marginal_unit}. Energy revenues therefore critically determine whether generators eventually operate at a loss and require make-whole payments to adhere to the dispatch.

Consequently, although the marginal unit pricing method could yield interpretable ancillary service prices, it cannot verify whether the derived prices are adequate, and thus is not appropriate for pricing voltage stability services.

\vspace{-0.3cm}
\subsection{Market Outcomes under Restricted Pricing Method}
The restricted pricing method preserves the binary nature of UC while producing shadow prices for unit commitment and market services, as described in Section~\ref{Restricted Pricing Method}. The resulting pricing outcomes are analyzed below.

\subsubsection{Ancillary Service and Energy Prices}\label{Prices restricted}

\begin{figure}[t]
\centering
\begin{adjustbox}{minipage=[t]{1\linewidth},center}
\begin{tikzpicture}
\begin{axis}[
    width=1\linewidth, height=3cm,
    xlabel={\scriptsize Time [h]},
     ylabel={\scriptsize \parbox[c]{2cm}{\centering Ancillary service prices [\texteuro/service]}}, 
    ylabel style={align=center, yshift=-0.1cm}, 
    tick label style={font=\scriptsize},
    xmin=1, xmax=24,
    xtick={4,8,12,16,20},
    ymin=0, ymax=30,
    ytick={0,10,20,30},
    grid=both, grid style={dashed, gray!60},
    legend style={
        font=\scriptsize,
        fill=none,
        draw=none,
        at={(0,1.05)}, anchor=north west,
        legend columns=2
    },
    legend cell align={left},
    legend image post style={scale=1.2, yshift=0cm}, 
    legend image code/.code={%
        \draw[mark repeat=1,mark phase=1,#1] plot coordinates {(0,0)}; 
    },
]
\addplot+[  mark=o, color=green!50!black, line width=0.6pt] coordinates {
(1,0.00016473594779495282)
(2,0.00016458381327772863)
(3,0.00016464547011034786)
(4,0.00016475540426811157)
(5,0.00016461701771032804)
(6,0.00016457026629677576)
(7,0)
(8,0)
(9,7.487540489670453)
(10,16.33976447079148)
(11,16.33976492009913)
(12,16.339763519769605)
(13,16.339763310936263)
(14,16.339763958612803)
(15,16.33976555732824)
(16,16.33976370337989)
(17,8.084000320749345)
(18,15.828854460588774)
(19,7.253382732428237)
(20,16.33976368961769)
(21,7.487532462140316)
(22,5.75716635108686)
(23,0.00016462401654711038)
(24,0.00016459543902441761)
};
\addlegendentry{$\Gamma_{{g_f}\textrm{-}b23}$}

\addplot+[  mark=star, color=red, line width=0.6pt] coordinates {
(1,0)
(2,0)
(3,0)
(4,0)
(5,0)
(6,0)
(7,0)
(8,0)
(9,11.685562398207065)
(10,25.500937895510013)
(11,25.500938558474164)
(12,25.50093647744107)
(13,25.500936151807533)
(14,25.500937111618036)
(15,25.50093934798408)
(16,25.500936744913265)
(17,12.279682207822091)
(18,24.044194907615932)
(19,11.017965495132707)
(20,25.500936759201597)
(21,11.685550211171561)
(22,9.324584222059658)
(23,0)
(24,0)
};
\addlegendentry{$\hat{Q}_{{g_f}\textrm{-}b23}$}

\addplot+[  mark=o, color=blue, line width=0.6pt] coordinates {
(1,0.00018466148785714674)
(2,0.00018449111658795035)
(3,0.00018456012184417644)
(4,0.00018468325478116437)
(5,0.00018452828538412933)
(6,0.0001844759480608735)
(7,0)
(8,0)
(9,0)
(10,0)
(11,-0)
(12,0)
(13,0)
(14,0)
(15,0)
(16,0)
(17,0)
(18,0)
(19,0)
(20,0)
(21,0)
(22,12.56333874108098)
(23,0.00018453611775228127)
(24,0.0001845041322145264)
};
\addlegendentry{$\Gamma_{{g_f}\textrm{-}b24}$}

\addplot+[  mark=star, color=orange, line width=0.6pt] coordinates {
(1,0)
(2,0)
(3,0)
(4,0)
(5,0)
(6,0)
(7,0)
(8,0)
(9,0)
(10,0)
(11,0)
(12,0)
(13,0)
(14,0)
(15,0)
(16,0)
(17,0)
(18,0)
(19,0)
(20,0)
(21,0)
(22,20.169827119478853)
(23,0)
(24,0)
};
\addlegendentry{$\hat{Q}_{{g_f}\textrm{-}b24}$}

\end{axis}
\end{tikzpicture}
\end{adjustbox}
\vspace{-0.9cm}
\caption{Shadow prices of $\hat{Q}_{g_f}$ [\texteuro/Mvar] and $\Gamma_{g_f}$ [\texteuro/MVA] for buses 23 and 24, calculated using restricted pricing method.} 
\label{fig:Shadow prices of service_restricted}
       \vspace{-0.8cm}
\end{figure}
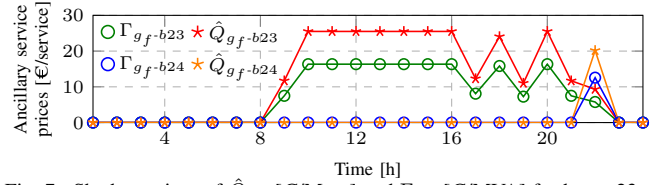

\begin{figure}[t]
\centering
\begin{adjustbox}{minipage=[t]{1\linewidth},center}
\begin{tikzpicture}

\begin{axis}[
    width=0.98\linewidth, height=3cm,
    xlabel={\scriptsize Time [h]},
    ylabel={\scriptsize \parbox[c]{3cm}{\centering Energy price\\[2pt]{[\texteuro/MWh]}}},
    ylabel style={align=center, yshift=-0.4cm},
    label style={font=\scriptsize}, 
    tick label style={font=\scriptsize},
    xmin=1, xmax=24,
    xtick={4,8,12,16,20},
    ymin=0, ymax=15,
    ytick distance=3,
    grid=both, grid style={dashed, gray!60},
    scaled y ticks = false,
    y tick label style = {
        /pgf/number format/fixed,
        /pgf/number format/precision=3
    },
    legend style={
        font=\scriptsize, 
        fill=none, 
        draw=none, 
        at={(0.143,0)}, anchor=south,
        legend columns=2
    },
    legend cell align={left},
    legend image post style={scale=1},
    legend image code/.code={%
        \draw[mark repeat=1,mark phase=1,#1] plot coordinates {(0,0)};
    },
    axis lines=box
]

\addplot+[color=red, solid, mark=o, mark size=1.5pt, line width=0.6pt]
coordinates {
(1,0)
(2,0)
(3,0)
(4,0)
(5,0)
(6,0)
(7,0)
(8,0)
(9,6.199998429328033)
(10,13.529999188778547)
(11,13.529999249568354)
(12,13.529998920966962)
(13,13.529998893048719)
(14,13.529998979913191)
(15,13.529999311129876)
(16,13.529998945366577)
(17,6.909988412735012)
(18,13.529998910570946)
(19,6.199995109288188)
(20,13.52999894857245)
(21,6.199997434652681)
(22,13.529999212369509)
(23,0)
(24,0)
};
\end{axis}
\end{tikzpicture}
\end{adjustbox}
\vspace{-1cm}
\caption{Hourly energy price over the whole horizon, calculated using restricted pricing method.}
\label{fig:hourly_energy_price}
\vspace{-0.6cm}
\end{figure}
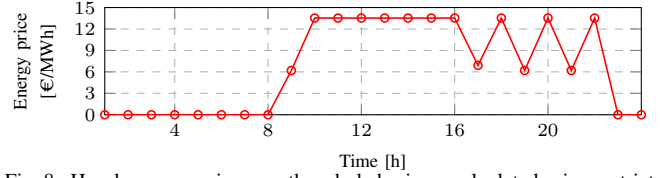

\begin{figure}[t]
\centering
\begin{adjustbox}{minipage=[t]{1\linewidth},center}
\begin{tikzpicture}
\begin{axis}[
    width=1\linewidth, height=3cm,
    xlabel={\scriptsize Time [h]},
    ylabel={\scriptsize \parbox[c]{3cm}{\centering Commitment price\\[1pt]{[k\texteuro/unit]}}}, 
    ylabel style={align=center, yshift=-0.3cm}, 
    label style={font=\scriptsize}, 
    tick label style={font=\scriptsize},
    xmin=1, xmax=24,
    ymin=0, ymax=2.8,
    ytick distance=0.70,
    xtick={4,8,12,16,20},
    grid=both, grid style={dashed, gray!60},
    scaled y ticks = false,
y tick label style={
    /pgf/number format/fixed,
    /pgf/number format/precision=3
},
    legend style={
        font=\scriptsize, 
        fill=none, 
        draw=none, 
        at={(0.143,0.18)}, anchor=south,
        legend columns=2
    },
    legend cell align={left},
    legend image post style={scale=1}, 
    legend image code/.code={%
        \draw[mark repeat=1,mark phase=1,#1] plot coordinates {(0,0)}; 
    },
    axis lines=box
]

\addplot+[  color=blue, solid, mark=star, mark size=1.5pt, line width=0.6pt]
coordinates {
(1,0)
(2,0)
(3,0)
(4,0)
(5,0)
(6,0)
(7,0)
(8,2.620484)
(9,0.11222)
(10,0)
(11,0)
(12,0)
(13,0)
(14,0)
(15,0)
(16,0)
(17,0.030942)
(18,0)
(19,0.119537)
(20,0)
(21,0.61222)
(22,0)
(23,0)
(24,0)
};
\addlegendentry{$g_c$-$b2$}

\addplot+[  color=red, solid, mark=o, mark size=1.5pt, line width=0.6pt]
coordinates {
(1,0.0)
(2,0.0)
(3,0.0)
(4,0.0)
(5,0.0)
(6,0.0)
(7,0.0)
(8,0.0)
(9,0.0)
(10,0.0)
(11,0.0)
(12,0.0)
(13,0.0)
(14,0.0)
(15,0.0)
(16,0.0)
(17,1.276979)
(18,0.0)
(19,0.367316)
(20,0.0)
(21,0.0)
(22,0.0)
(23,0.0)
(24,0.0)
};
\addlegendentry{$g_c$-$b3$}

\addplot+[  color=green!50!black, solid, mark=o, mark size=1.5pt, line width=0.6pt]
coordinates {
(1,0.315188)
(2,0.315188)
(3,0.315188)
(4,0.315189)
(5,0.315188)
(6,0.315188)
(7,0.315210)
(8,0.315260)
(9,0.0)
(10,0.0)
(11,0.0)
(12,0.0)
(13,0.0)
(14,0.0)
(15,0.0)
(16,0.0)
(17,0.0)
(18,0.0)
(19,0.0)
(20,0.0)
(21,0.0)
(22,0.0)
(23,0.315188)
(24,0.315188)
};
\addlegendentry{$g_c$-$b27$}

\end{axis}

\end{tikzpicture}
\end{adjustbox}
\vspace{-1cm}
\caption{Commitment prices captured by SGs at buses \{2, 3, 27\}, calculated by \eqref{eq:SG_capture_commit_price} at each hour. The price for other SGs is zero as they stay offline over the whole horizon.}
\label{fig:UC_price}
        \vspace{-0.5cm}
\end{figure}
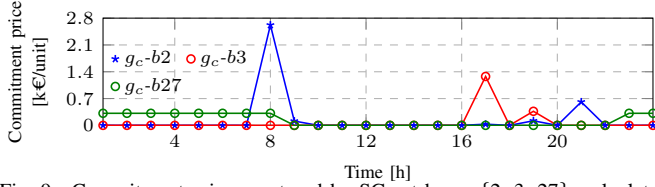

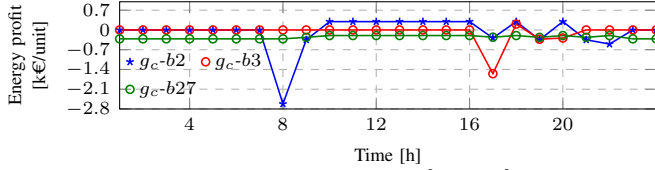
\begin{figure}[t]
\centering
\begin{adjustbox}{minipage=[t]{1\linewidth},center}
\begin{tikzpicture}
\begin{axis}[
    width=0.98\linewidth, height=3cm,
    xlabel={\scriptsize Time [h]},
    ylabel={\scriptsize \parbox[c]{2cm}{\centering Energy profit\\[1pt]{[k\texteuro/unit]}}}, 
    ylabel style={align=center, yshift=-0.4cm}, 
    label style={font=\scriptsize}, 
    tick label style={font=\scriptsize},
    xmin=1, xmax=24,
    xtick={4,8,12,16,20},
    ymin=-2.8, ymax=1,
    ytick distance=0.7,
    grid=both, grid style={dashed, gray!60},
    scaled y ticks = false,
y tick label style = {
    /pgf/number format/fixed,
    /pgf/number format/precision=3
},
    legend style={
        font=\scriptsize, 
        fill=none, 
        draw=none, 
        at={(0.143,0)}, anchor=south,
        legend columns=2
    },
    legend cell align={left},
    legend image post style={scale=1}, 
    legend image code/.code={%
        \draw[mark repeat=1,mark phase=1,#1] plot coordinates {(0,0)}; 
    },
    axis lines=box
]

\addplot+[  color=blue, solid, mark=star, mark size=1.5pt, line width=0.6pt]
coordinates {
(1,0)
(2,0)
(3,0)
(4,0)
(5,0)
(6,0)
(7,0)
(8,-2.620573)
(9,-0.3486)
(10,0.294974)
(11,0.294974)
(12,0.294974)
(13,0.294974)
(14,0.294974)
(15,0.294974)
(16,0.294974)
(17,-0.286263)
(18,0.294974)
(19,-0.3486)
(20,0.294974)
(21,-0.3486)
(22,-0.5)
(23,0)
(24,0)
};
\addlegendentry{$g_c$-$b2$}

\addplot+[  color=red, solid, mark=o, mark size=1.5pt, line width=0.6pt]
coordinates {
(1,0)
(2,0)
(3,0)
(4,0)
(5,0)
(6,0)
(7,0)
(8,0)
(9,0)
(10,0)
(11,0)
(12,0)
(13,0)
(14,0)
(15,0)
(16,0)
(17,-1.550201)
(18,0.208216)
(19,-0.327464)
(20,-0.285)
(21,0)
(22,0)
(23,0)
(24,0)
};
\addlegendentry{$g_c$-$b3$}

\addplot+[  color=green!50!black, solid, mark=o, mark size=1.5pt, line width=0.6pt]
coordinates {
(1,-0.31526)
(2,-0.31526)
(3,-0.31526)
(4,-0.31526)
(5,-0.31526)
(6,-0.31526)
(7,-0.31526)
(8,-0.31526)
(9,-0.261527)
(10,-0.198)
(11,-0.198)
(12,-0.198)
(13,-0.198)
(14,-0.198)
(15,-0.198)
(16,-0.198)
(17,-0.255373)
(18,-0.198)
(19,-0.261527)
(20,-0.198)
(21,-0.261527)
(22,-0.198)
(23,-0.31526)
(24,-0.31526)
};
\addlegendentry{$g_c$-$b27$}

\end{axis}
\end{tikzpicture}
\end{adjustbox}
\vspace{-0.9cm}
\caption{Hourly energy profit of SGs at buses \{2, 3, 27\}, calculated using restricted pricing method. The energy profit of other SGs is zero as they stay offline over the whole horizon.}
\label{fig:SG_energy_profit_restricted}
        \vspace{-0.4cm}
\end{figure}

\begin{table}[t]
\caption{Market Outcome Using Restricted Pricing Method (k\texteuro/Day) }
         \vspace{-0.3cm}
\centering
\setlength{\tabcolsep}{1.5pt}
{\fontsize{8pt}{12pt}\selectfont
\begin{tabular}{ccccccc}
\toprule
\makecell[c]{ \\ Units \\ } &
\makecell[c]{ \\ Energy \\ profits } &
\makecell[c]{ \\ Remuneration by \\ commitment} &
\multicolumn{2}{c}{\makecell{Remuneration by\\voltage stability services}} &
\makecell[c]{ \\ Total \\  profit } \\
\cmidrule(lr){4-5}
& & &  $\hat{Q}_{g_f}$ & $\mathrm{SCR}_{\Phi(g_f)}$ \\
\midrule
$g_v$-$b1$    & $\mathrm{11.26}$  & $\mathrm{0}$    & $\mathrm{0}$ & $\mathrm{6.17}$  & $\mathrm{17.42}$  \\ \hline
$g_c$-$b2$    & $\mathrm{-1.80}$  & $\mathrm{3.50}$  & $\mathrm{0}$ & $\mathrm{4.17}$ & $\mathrm{5.87}$ \\ 
$g_c$-$b3$    & $\mathrm{-1.96}$  & $\mathrm{1.64}$  & $\mathrm{0}$ & $\mathrm{0.65}$ & $\mathrm{0.34}$ \\ 
$g_c$-$b27$   & $\mathrm{-6.17}$  & $\mathrm{3.15}$  & $\mathrm{0}$ & $\mathrm{9.32}$ & $\mathrm{6.30}$ \\ 
other SGs     & $\mathrm{0}$       & $\mathrm{0}$       & $\mathrm{0}$ & $\mathrm{0}$     & $\mathrm{0}$ \\ \hline
$g_f$-$b23$   & $\mathrm{10.03}$  & $\mathrm{0}$    & $\mathrm{14.50}$ & $\mathrm{0}$ & $\mathrm{24.53}$ \\ 
$g_f$-$b24$   & $\mathrm{19.16}$  & $\mathrm{0}$    & $\mathrm{16.56}$ & $\mathrm{0}$ & $\mathrm{35.72}$ \\
\bottomrule
\end{tabular}
}
\label{tab:unit_revenue_restricted}
         \vspace{-0.4cm}
\end{table}

The prices of static voltage stability services are shown in Fig.~\ref{fig:Shadow prices of service_restricted}. Before 08:00, the system operates under light-load conditions, and only a part of RES generation is needed to meet electricity demand without triggering the voltage stability constraints, resulting in zero ancillary service prices. It is noteworthy that the service prices for bus 24 are non-zero only at 22:00. This indicates that bus 24 does not procure the ancillary service during periods when the voltage stability constraint at bus 23 is binding, since the needed SCR and reactive power support for bus 24 have already been provided through satisfying the constraint on bus 23. This is consistent with the fact that the SCR of bus 24 is typically higher (as illustrated in Fig.~\ref{fig:UC_strength}) and that bus 24 can absorb the reactive injection of $g_f$–$b23$. Furthermore, the price of $\hat{Q}_{g_f}$ is always higher than that of $\Gamma_{g_f}$, implying that reactive power support generally plays a more critical role than SCR enhancement in securing static voltage stability (as defined in \eqref{eq:explanation_voltage_stability}) at GFL-IBR buses. 

As can be seen from the energy prices in Fig.~\ref{fig:hourly_energy_price}, the voltage stability service prices exhibit a similar trend, implying that RES and SGs at buses \{2,3,27\} (as discussed in Section~\ref{Remuneration restricted}) are alternatively chosen to clear markets in this case. Such consistency arises because the restricted pricing method identifies units for clearing both energy and ancillary service markets by fixing the optimal UC schedule. This makes the pricing process align closely with the strict definition of marginal cost and effectively capture the coupling across different markets.

\subsubsection{Remuneration to Generation Units}\label{Remuneration restricted}
As discussed in Section~\ref{Restricted Pricing Method}, VSGs and GFL-IBR, which do not have commitment variables, cannot be remunerated through $\lambda_{g_c,\mathrm{commit}}$. As depicted in Fig.~\ref{fig:UC_price}, only the SGs at buses \{2, 3, 27\} are committed and thus eligible to receive commitment payments. The price for $g_c\textrm{-}b2$ spikes at 08:00 due to unit startup and the associated fixed costs, while the resulting negative energy profit (defined as energy revenue minus operating cost, as in Fig.~\ref{fig:SG_energy_profit_restricted}) is largely offset by this payment. A similar case is observed for $g_c\textrm{-}b3$ at 17:00.

According to Eqs.~\eqref{eq:SG_capture_commit_price} and~\eqref{eq:profit_units}, the profits of all units under the market clearing prices derived from the restricted pricing method are further calculated and summarized in Table~\ref{tab:unit_revenue_restricted}. Since RES are modeled without explicit operating costs, both VSGs and GFL-IBR achieve substantial profits in the market. By contrast, the energy revenues of thermal units are insufficient to fully recover their operating costs; however, revenues from ancillary services and commitment payments make up the shortfall, yielding positive net profits overall. In particular, unit $g_c\textrm{-}b27$ is electrically close to buses 23 and 24, where voltage stability support is required. As a result, it can provide significant SCR enhancement and therefore earns substantially higher ancillary service revenues than the other thermal units.

Notably, unit $g_c\textrm{-}b3$ exhibits the lowest profit. Its deficit in energy profit can only be offset by the combined remuneration from ancillary services and commitment prices, rather than either incentive alone, to guarantee its revenue adequacy. However, with appropriately defined uplift payments for compensation, these prices could be used as the market prices, a capability realized by the restricted pricing method.

\subsubsection{Impact of GFL-IBR Reactive Capacity on Ancillary Service Prices and SG Net Profits}

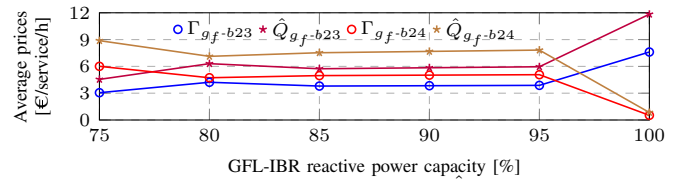
\begin{figure}[t]
\centering
\begin{adjustbox}{minipage=[t]{1\linewidth},center}
\begin{tikzpicture}

\begin{axis}[
    width=1\linewidth, height=3cm,
    xlabel={\scriptsize GFL-IBR reactive power capacity [\%]},
    ylabel={\scriptsize \parbox[c]{2cm}{\centering Average prices [\texteuro/service/h]}}, 
    ylabel style={align=center, yshift=-0.3cm},
    label style={font=\scriptsize}, 
    tick label style={font=\scriptsize},
    xmin=75, xmax=100,
    ymin=0, ymax=12,
    xtick={75,80,85,90,95,100},
    ytick={0,3,6,9,12},
    grid=both, 
    grid style={dashed, gray!60},
    legend style={
        font=\scriptsize,
        fill=none,
        draw=none,
        at={(0.12,1.05)}, anchor=north west,
        legend columns=4,
         row sep=-4pt 
    },
    legend cell align={left},
    legend image post style={scale=1},
    legend image code/.code={%
        \draw[mark repeat=1,mark phase=1,#1] plot coordinates {(0,0)};
    },
    axis lines=box
]

\addplot+[ color=blue, solid, mark=o, mark size=1.5pt, line width=0.6pt]
coordinates {
     (75, 3.055198147478158) 
     (80, 4.222369067247352)
     (85, 3.7988792523733905)
     (90, 3.837390956785262)
     (95, 3.8758726596792052)
     (100, 7.609072636853163 )
};
\addlegendentry{$\Gamma_{g_f\textrm{-}b23}$}

\addplot+[  color=purple, solid, mark=star, mark size=1.5pt, line width=0.6pt]
coordinates {
     (75, 4.54644691141595)
     (80, 6.304930749966811)
     (85, 5.735180319691235)
     (90, 5.847966128393447)
     (95, 5.958623151219072)
     (100, 11.835209939112113 )
};
\addlegendentry{$\hat{Q}_{g_f\textrm{-}b23}$}

\addplot+[  color=red, solid, mark=o, mark size=1.5pt, line width=0.6pt]
coordinates {
     (75, 6.0055324137927615) 
     (80, 4.729713004010225)
     (85, 4.968942282587732)
     (90, 5.020143819304367)
     (95, 5.071272899163448)
     (100, 0.5235258639876647 )
};
\addlegendentry{$\Gamma_{g_f\textrm{-}b24}$}

\addplot+[  color=brown, solid, mark=star, mark size=1.5pt, line width=0.6pt]
coordinates {
     (75, 8.879547539220349 )
     (80, 7.123573782127177 )
     (85, 7.528107555259125 )
     (90, 7.677169483700452 )
     (95, 7.823386971458903 )
     (100, 0.8404095532841933 )
};
\addlegendentry{$\hat{Q}_{g_f\textrm{-}b24}$}

\end{axis}

\end{tikzpicture}
\vspace{-2cm}
\end{adjustbox}

\vspace{-0.5cm}
\caption{Average value of ancillary service prices ($\hat{Q}_{g_f}$ [\texteuro/Mvar/h] and $\Gamma_{g_f}$ [\texteuro/MVA/h]) over 24-h operation using restricted pricing method, as a function of variable GFL-IBR reactive power capacity in the system.} 
\label{fig:mean_prices_restricted}
        \vspace{-0.5cm}
\end{figure}

\begin{figure}[t]
\centering
\begin{adjustbox}{minipage=[t]{1\linewidth},center}
\begin{tikzpicture}

\begin{axis}[
    width=0.98\linewidth, height=3cm,
    xlabel={\scriptsize GFL-IBR reactive power capacity [\%]},
    ylabel={\scriptsize \parbox[c]{2cm}{\centering Net profit [\texteuro/unit/h]}}, 
    ylabel style={align=center, yshift=-0.4cm},
    label style={font=\scriptsize}, 
    tick label style={font=\scriptsize},
    xmin=75, xmax=100,
    xtick={75,80,85,90,95,100},
    ymin=0, ymax=420,
    ytick distance=105,
    grid=both, grid style={dashed, gray!60},
    scaled y ticks = false,
    y tick label style = {
        /pgf/number format/fixed,
        /pgf/number format/precision=1
    },
    legend style={
        font=\scriptsize, 
        fill=none, 
        draw=none, 
        at={(0.143,0)}, anchor=south,
        legend columns=2
    },
    legend cell align={left},
    legend image post style={scale=1},
    legend image code/.code={%
        \draw[mark repeat=1,mark phase=1,#1] plot coordinates {(0,0)};
    },
    axis lines=box
]

\addplot+[color=blue, solid, mark=o, mark size=1.5pt, line width=0.6pt]
coordinates {
(75,238.04492325246187)
(80,233.57085972105543)
(85,275.9240625907716)
(90,277.68968819914267)
(95,279.44564552448827)
(100,244.46622055547502)
};
\addlegendentry{$g_c\textrm{-}b2$}

\addplot+[color=red, solid, mark=o, mark size=1.5pt, line width=0.6pt]
coordinates {
(75,12.549804252900534)
(80,5.515484356913807)
(85,13.42603377632537)
(90,13.55008724289729)
(95,13.669844889522944)
(100,14.243655537861814)
};
\addlegendentry{$g_c\textrm{-}b3$}

\addplot+[color=green!50!black, solid, mark=o, mark size=1.5pt, line width=0.6pt]
coordinates {
(75,394.19044239501386)
(80,364.61736141865254)
(85,368.9385659170105)
(90,374.1384461748985)
(95,379.33318461479024)
(100,262.64482524183484)
};
\addlegendentry{$g_c\textrm{-}b27$}

\end{axis}
\end{tikzpicture}
\end{adjustbox}
\vspace{-0.9cm}
\caption{Average net profits of SGs at buses \{2,3,27\} over the horizon, calculated by the restricted pricing method, as a function of variable GFL-IBR reactive power capacity in the system. The profits of the remaining SGs are zero since they are not dispatched.}
\label{fig:restricted_net_profit}
\vspace{-0.4cm}
\end{figure}
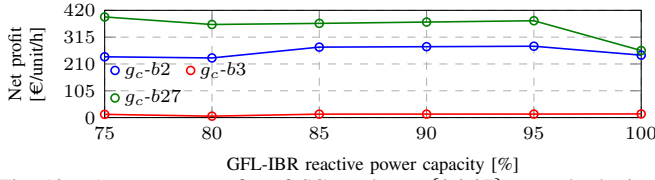

It is worth noting that the ancillary service prices are also closely associated with the reactive power support of GFL-IBR and may show opposing variation trends for different buses, as depicted in Fig.~\ref{fig:mean_prices_restricted}. When the permissible reactive power output is below $95\%$, the service prices for bus 24 are generally higher than those for bus 23. This is because, to enable $g_f\textrm{-}b24$ to deliver greater wind power (as shown in Fig.~\ref{fig:wind_curt_and_cost}) into the grid without violating voltage stability constraints, bus 24 becomes more dependent on ancillary services. Once the penetration hits the $95\%$ threshold, the prices for bus 24 drop, while a rising trend for prices is seen for bus 23. This change occurs since the system SCR stays fixed and the voltage stability is secured primarily through elevated reactive power injections. Since bus 23 features a generally lower SCR (as in Fig.~\ref{fig:UC_strength}), its voltage stability constraint tends to bind when the active injection from $g_f\textrm{-}b23$ is needed, driving up its ancillary service prices. In contrast, bus 24 with higher SCR can benefit from bus 23, ensuring the voltage stability with lower or zero service prices. 

Further insights can be obtained by relating these price trends to the results in Fig.~\ref{fig:wind_curt_and_cost}. Although expanding the reactive power capability of GFL-IBR improves wind power utilization and dispatch economy, the ancillary service prices remain nearly flat over the $80\%$-$95\%$ reactive power capacity range. Within the apparent power limit in~\eqref{eq:cons_IBR_capacity}, additional wind power can be injected while maintaining greater reactive power headroom, without requiring additional marginal reactive power support for voltage stability. Beyond $95\%$ capacity, the ancillary service prices at buses 23 and 24 vary according to the power interactions between the two buses and their respective SCR levels. In comparison, the price curves under dispatchable pricing in Fig.~\ref{fig:mean_prices_dispatchable} exhibit greater sensitivity to the available reactive power capacity. This behavior stems from the relaxation of binary variables in the dispatchable formulation, which alters the evaluation of marginal contributions in the pricing process.

The unit net profits, as shown in Fig. \ref{fig:restricted_net_profit}, vary with reactive power injection levels. Similar to the results in Table \ref{tab:unit_revenue_restricted}, $g_c\textrm{-}b27$ consistently earns much more from ancillary services than other thermal units, leading to substantial net profits. Even with very limited profitability, $g_c\textrm{-}b3$ would still remain dispatchable, which demonstrates that commitment prices under the restricted pricing method could allow derived service prices to robustly support that no units have incentives to leave the market.

\subsection{Market Outcomes under Dispatchable Pricing Method}\label{Market Outcomes under Dispatchable Pricing Method}
As outlined in Section~\ref{Dispatchable Pricing Method}, binary variables are relaxed to compute shadow prices under the dispatchable pricing approach, converting the MISOCP \eqref{eq:MISOCP} into an SOCP. The market clearing prices and remuneration to generators under this approach are analyzed as follows.

\subsubsection{Ancillary Service and Energy Prices}\label{Prices dispatchable}

\begin{figure}[t]
\centering
\begin{adjustbox}{minipage=[t]{1\linewidth},center}
\begin{tikzpicture}
\begin{axis}[
    width=1\linewidth, height=3cm,
    xlabel={\scriptsize Time [h]},
     ylabel={\scriptsize \parbox[c]{2cm}{\centering Ancillary service prices [\texteuro/service]}}, 
    ylabel style={align=center, yshift=-0.1cm},
    tick label style={font=\scriptsize},
    xmin=1, xmax=24,
    xtick={4,8,12,16,20},
    ymin=0, ymax=30,
    ytick={0,10,20,30},
    grid=both, grid style={dashed, gray!60},
    legend style={
        font=\scriptsize,
        fill=none,
        draw=none,
        at={(0,1.09)}, anchor=north west,
        legend columns=4
    },
    legend cell align={left},
    legend image post style={scale=1.2, yshift=0cm},
    legend image code/.code={%
        \draw[mark repeat=1,mark phase=1,#1] plot coordinates {(0,0)};
    },
]
\addplot+[  mark=o, color=green!50!black, line width=0.6pt] coordinates {
(1,2.9756181167409035)
(2,0)
(3,0)
(4,6.245751598157395)
(5,0)
(6,0)
(7,2.328639246863979)
(8,2.192923028016749)
(9,2.1683114975265765)
(10,13.258172399024641)
(11,14.434385765744203)
(12,8.122150201418128)
(13,8.122163036862169)
(14,8.122181712914914)
(15,14.295790305817647)
(16,8.085880202515678)
(17,15.089753770354784)
(18,6.498691441731704)
(19,12.191633327157906)
(20,1.6834109380381916)
(21,2.3787894381272903)
(22,2.9494078391828755)
(23,2.197311446141071)
(24,1.3724405658499919)
};
\addlegendentry{$\Gamma_{{g_f}\textrm{-}b23}$}

\addplot+[  mark=star, color=red, line width=0.6pt] coordinates {
(1,5.1146279867909925)
(2,0)
(3,0)
(4,10.735368866221467)
(5,0)
(6,0)
(7,3.8270068703539453)
(8,3.4687507600191285)
(9,3.3790051264033707)
(10,20.226251942249178)
(11,22.02063749364638)
(12,12.390951862397415)
(13,12.390969694311828)
(14,12.391000178985703)
(15,21.720764797044353)
(16,12.285602098006603)
(17,22.853396057732873)
(18,9.842318110266604)
(19,18.46431611496414)
(20,2.582846274328699)
(21,3.7542507007095027)
(22,4.790241396171887)
(23,3.8117887317133023)
(24,2.396504343270279)
};
\addlegendentry{$\hat{Q}_{{g_f}\textrm{-}b23}$}

\addplot+[  mark=o, color=blue, line width=0.6pt] coordinates {
(1,6.564417559499114)
(2,0)
(3,0)
(4,13.764111244781523)
(5,0)
(6,0)
(7,5.124024570588341)
(8,7.317170730751714)
(9,7.557868721614166)
(10,0)
(11,0)
(12,0)
(13,0)
(14,0)
(15,0)
(16,0)
(17,0)
(18,0)
(19,0)
(20,8.14087068291732)
(21,7.087081219074674)
(22,6.799093540086149)
(23,4.786959972184416)
(24,2.967009119603834)
};
\addlegendentry{$\Gamma_{{g_f}\textrm{-}b24}$}

\addplot+[  mark=star, color=orange, line width=0.6pt] coordinates {
(1,11.22848183498444)
(2,0)
(3,0)
(4,23.543596693255168)
(5,0)
(6,0)
(7,8.355786769400666)
(8,11.503812065827798)
(9,11.716308964536193)
(10,0)
(11,0)
(12,0)
(13,0)
(14,0)
(15,0)
(16,0)
(17,0)
(18,0)
(19,0)
(20,12.436712756602102)
(21,11.118284415742455)
(22,10.959531100514665)
(23,8.273032760861675)
(24,5.166062370466849)
};
\addlegendentry{$\hat{Q}_{{g_f}\textrm{-}b24}$}

\end{axis}
\end{tikzpicture}
\end{adjustbox}
\vspace{-0.9cm}
\caption{Shadow prices of $\hat{Q}_{g_f}$ [\texteuro/Mvar] and $\Gamma_{g_f}$ [\texteuro/MVA] for buses 23 and 24, calculated using dispatchable pricing method.} 
\label{fig:Shadow prices of service_dispatchable}
\vspace{-1cm}
\end{figure}
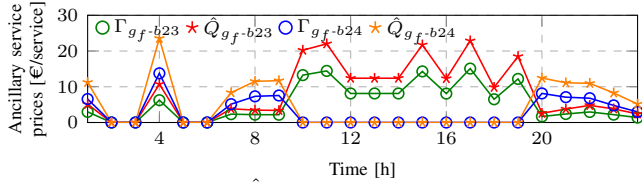

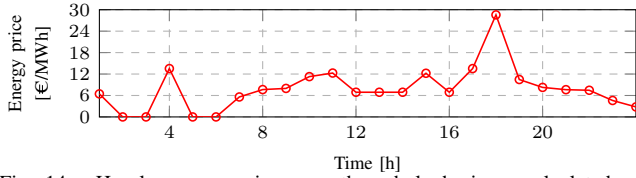
\begin{figure}[!t]
\centering
\begin{adjustbox}{minipage=[t]{1\linewidth},center}
\begin{tikzpicture}

\begin{axis}[
    width=0.98\linewidth, height=3cm,
    xlabel={\scriptsize Time [h]},
    ylabel={\scriptsize \parbox[c]{3cm}{\centering Energy price\\[2pt]{[\texteuro/MWh]}}},
    ylabel style={align=center, yshift=-0.4cm},
    label style={font=\scriptsize}, 
    tick label style={font=\scriptsize},
    xmin=1, xmax=24,
    xtick={4,8,12,16,20},
    ymin=0, ymax=30,
    ytick distance=6,
    grid=both, grid style={dashed, gray!60},
    scaled y ticks = false,
    y tick label style = {
        /pgf/number format/fixed,
        /pgf/number format/precision=3
    },
    legend style={
        font=\scriptsize, 
        fill=none, 
        draw=none, 
        at={(0.143,0)}, anchor=south,
        legend columns=2
    },
    legend cell align={left},
    legend image post style={scale=1},
    legend image code/.code={%
        \draw[mark repeat=1,mark phase=1,#1] plot coordinates {(0,0)};
    },
    axis lines=box
]

\addplot+[color=red, solid, mark=o, mark size=1.5pt, line width=0.6pt]
coordinates {
(1,6.449917504533006)
(2,0)
(3,0)
(4,13.52993375956526)
(5,0)
(6,0)
(7,5.5872765448921315)
(8,7.654365125080791)
(9,7.978555369298991)
(10,11.279106381199119)
(11,12.279755701199171)
(12,6.90962084989625)
(13,6.909623950393786)
(14,6.909629191567967)
(15,12.216298335854516)
(16,6.909513817516958)
(17,13.529166302713127)
(18,28.555365555917028)
(19,10.468668179268716)
(20,8.26229634048801)
(21,7.627344495833373)
(22,7.459066763028938)
(23,4.601012294718174)
(24,2.801724194294596)
};
\end{axis}
\end{tikzpicture}
\end{adjustbox}
\vspace{-1cm}
\caption{Hourly energy price over the whole horizon, calculated using dispatchable pricing method.}
\label{fig:hourly_energy_price_dispatchable}
\vspace{-0.5cm}
\end{figure}

The ancillary service prices are given in Fig.~\ref{fig:Shadow prices of service_dispatchable}, indicating that the price of $\hat{Q}_{g_f}$ remains higher than that of $\Gamma_{g_f}$, consistent with the observation in the restricted method. As analyzed in Section \ref{Prices restricted} and Fig. \ref{fig:Shadow prices of service_restricted}, voltage stability constraints are not binding before 08:00, implying that no ancillary services are required during this period. However, the prices produced by the dispatchable method contradict this fact, as non-zero price signals appear in several earlier time intervals, particularly at 04:00. Meanwhile, the variation pattern of ancillary service prices no longer closely matches that of energy prices, particularly at 18:00 in Fig.~\ref{fig:hourly_energy_price_dispatchable}, and differs noticeably from the pricing outcomes under the restricted method.

The comparison above suggests that relaxing binary commitment variables treats the startup/shutdown and no-load costs of SGs as variable rather than fixed in their bids. This enables such costs to be recovered via energy and ancillary service prices and potentially eliminates make-whole requirements. Conversely, this relaxation would lead to misrepresentation of ancillary service provision and spurious price signals, as it implicitly assumes that SGs connect only part of their self-impedance to the grid for pricing purposes, resulting in prices that tend to capture only partial voltage stability contributions from SGs (i.e., SCR enhancement). Furthermore, as an outer-approximation technique, McCormick envelopes used to linearize product terms generally expand the feasible region of continuous problems (i.e., dispatchable pricing method) and reduce constraint tightness, which may distort shadow prices.

Due to the results above, the dispatchable method would compromise the exact coupling between energy and ancillary service markets, and the resulting shadow prices would not precisely capture the contributions of marginal units to the system. Consequently, these prices may create issues. First, such prices would cause wealth transfers across market participants decoupled from actual system security needs. Generators may receive inflated or suppressed remuneration for services, which does not reflect their genuine physical contribution under the original mixed‑integer UC formulation. Second, long‑term investment may be misguided. If adopted as public reference prices for SCR ancillary services, they could drive sub‑optimal siting and capacity decisions: over‑investment at nodes with spurious high prices and insufficient investment at locations with real voltage stability requirements. These findings demonstrate the necessity of maintaining model integrality for releasing efficient price signals.

\subsubsection{Remuneration to Generation Units}\label{remuneration dispatchable}

\begin{table}[!t]
\caption{Market Outcome Using Dispatchable Pricing Method (k\texteuro/Day) }
         \vspace{-0.3cm}
\centering
\setlength{\tabcolsep}{9pt}
{\fontsize{8pt}{12pt}\selectfont
\begin{tabular}{cccccc}
\toprule
\makecell[c]{ \\ Units \\ } &
\makecell[c]{ \\ Energy \\ profits } &
\multicolumn{2}{c}{\makecell{Remuneration by\\voltage stability services}} &
\makecell[c]{ \\ Total \\  profit } \\
\cmidrule(lr){3-4}
& &   $\hat{Q}_{g_f}$ & $\mathrm{SCR}_{\Phi(g_f)}$  \\
\midrule
$g_v$-$b1$    & $\mathrm{13.05}$  & $\mathrm{0}$ & $\mathrm{7.69}$  & $\mathrm{20.74}$  \\ \hline
$g_c$-$b2$    & $\mathrm{-2.20}$  & $\mathrm{0}$ & $\mathrm{3.58}$ & $\mathrm{1.38}$ \\ 
$g_c$-$b3$    & $\mathrm{-0.25}$  & $\mathrm{0}$ & $\mathrm{0.71}$ & $\mathrm{0.46}$ \\ 
$g_c$-$b27$   & $\mathrm{-6.17}$  & $\mathrm{0}$ & $\mathrm{11.29}$ & $\mathrm{5.11}$ \\ 
other SGs     & $\mathrm{0}$       & $\mathrm{0}$ & $\mathrm{0}$ & $\mathrm{0}$ \\ \hline
$g_f$-$b23$   & $\mathrm{11.28}$  & $\mathrm{14.41}$ & $\mathrm{0}$ & $\mathrm{25.69}$ \\ 
$g_f$-$b24$   & $\mathrm{21.43}$  & $\mathrm{18.83}$ & $\mathrm{0}$ & $\mathrm{40.25}$ \\
\bottomrule
\end{tabular}
}
\label{tab:unit_revenue_dispatch}
         \vspace{-0.5cm}
\end{table}

The profits obtained by units are summarized in Table~\ref{tab:unit_revenue_dispatch}. Clearly, revenue generated from voltage stability services alone is able to cover the energy profit shortfall of thermal units, offering sufficient incentives for these units to stay operational in the market. Similar to the findings in Table~\ref{tab:unit_revenue_restricted}, $g_c\textrm{-}b27$ earns noticeably higher ancillary service revenues than other thermal generators, owing to its substantial SCR contribution to the relevant buses. Such revenue bridges its considerable profit gap and enables it to achieve the largest net profit among all thermal units.

It should be emphasized that the positive‑profit results shown in Table~\ref{tab:unit_revenue_restricted} and Table~\ref{tab:unit_revenue_dispatch} verify revenue adequacy for dispatched units under the two pricing schemes. This check does not, however, quantify lost opportunity costs. Full quantitative comparison of opportunity‑cost metrics between restricted and dispatchable pricing is reserved for future investigation, as it enables further examination of conditions for achieving a competitive equilibrium point.

\subsubsection{Impact of GFL-IBR Reactive Capacity on Ancillary Service Prices and SG Net Profits}

\begin{figure}[!t]
\centering
\begin{adjustbox}{minipage=[t]{1\linewidth},center}
\begin{tikzpicture}

\begin{axis}[
    width=1\linewidth, height=3cm,
    xlabel={\scriptsize GFL-IBR reactive power capacity [\%]},
    ylabel={\scriptsize \parbox[c]{2cm}{\centering Average prices [\texteuro/service/h]}}, 
    ylabel style={align=center, yshift=-0.3cm},
    label style={font=\scriptsize}, 
    tick label style={font=\scriptsize},
    xmin=75, xmax=100,
    ymin=0, ymax=14,
    xtick={75,80,85,90,95,100},
    ytick={0,2,4,6,8,10,12,14},
    grid=both, 
    grid style={dashed, gray!60},
    legend style={
        font=\scriptsize,
        fill=none,
        draw=none,
        at={(0,0.65)}, anchor=north west,
        legend columns=2,
         row sep=-4pt 
    },
    legend cell align={left},
    legend image post style={scale=1},
    legend image code/.code={%
        \draw[mark repeat=1,mark phase=1,#1] plot coordinates {(0,0)};
    },
    axis lines=box
]

\addplot+[ color=blue, solid, mark=o, mark size=1.5pt, line width=0.6pt]
coordinates {
     (75, 0.8427057885011346) 
     (80, 0.8677919344372462)
     (85, 0.9405705199196429)
     (90, 2.6245482994653297)
     (95, 3.495397909203067)
     (100, 5.630362544245243)
};
\addlegendentry{$\Gamma_{g_f\textrm{-}b23}$}

\addplot+[  color=purple, solid, mark=star, mark size=1.5pt, line width=0.6pt]
coordinates {
     (75, 1.3447460914721108)
     (80, 1.4024218498229557)
     (85, 1.5320726304660142)
     (90, 4.042834273219867)
     (95, 5.3972601144014485)
     (100, 8.7124326870538)
};
\addlegendentry{$\hat{Q}_{g_f\textrm{-}b23}$}

\addplot+[  color=red, solid, mark=o, mark size=1.5pt, line width=0.6pt]
coordinates {
     (75, 8.209737668731838) 
     (80, 8.241495338504336)
     (85, 7.698397046846825)
     (90, 5.8381031759071815)
     (95, 5.027639304788816)
     (100, 2.897806827863132)
};
\addlegendentry{$\Gamma_{g_f\textrm{-}b24}$}

\addplot+[  color=brown, solid, mark=star, mark size=1.5pt, line width=0.6pt]
coordinates {
     (75, 12.060348730838)
     (80, 12.252816219991749)
     (85, 11.636676340713494)
     (90, 9.0313754906864)
     (95, 7.897631469797052)
     (100, 4.725963315567751)
};
\addlegendentry{$\hat{Q}_{g_f\textrm{-}b24}$}

\end{axis}

\end{tikzpicture}
\end{adjustbox}

\vspace{-0.48cm}
\caption{Average value of ancillary service prices ($\hat{Q}_{g_f}$ [\texteuro/Mvar/h] and $\Gamma_{g_f}$ [\texteuro/MVA/h]) over 24-h operation using dispatchable pricing method, as a function of variable GFL-IBR reactive power capacity in the system.} 
\label{fig:mean_prices_dispatchable}
       \vspace{-0.5cm}
\end{figure}
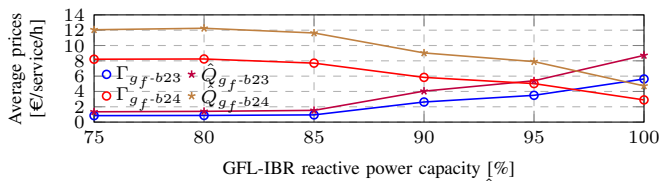

\begin{figure}[!t]
\centering
\begin{adjustbox}{minipage=[t]{1\linewidth},center}
\begin{tikzpicture}

\begin{axis}[
    width=0.98\linewidth, height=3cm,
    xlabel={\scriptsize GFL-IBR reactive power capacity [\%]},
    ylabel={\scriptsize \parbox[c]{2cm}{\centering Net profit [\texteuro/unit/h]}}, 
    ylabel style={align=center, yshift=-0.4cm},
    label style={font=\scriptsize}, 
    tick label style={font=\scriptsize},
    xmin=75, xmax=100,
    xtick={75,80,85,90,95,100},
    ymin=0, ymax=350,
    ytick distance=70,
    grid=both, grid style={dashed, gray!60},
    scaled y ticks = false,
    y tick label style = {
        /pgf/number format/fixed,
        /pgf/number format/precision=1
    },
    legend style={
        font=\scriptsize, 
        fill=none, 
        draw=none, 
        at={(0.243,0.2)}, anchor=south,
        legend columns=2
    },
    legend cell align={left},
    legend image post style={scale=1},
    legend image code/.code={%
        \draw[mark repeat=1,mark phase=1,#1] plot coordinates {(0,0)};
    },
    axis lines=box
]

\addplot+[color=blue, solid, mark=o, mark size=1.5pt, line width=0.6pt]
coordinates {
(75,77.85552288854826)
(80,76.48696461741548)
(85,78.10557984812863)
(90,66.91528212951935)
(95,61.89887532273335)
(100,55.94826661811305)
};
\addlegendentry{$g_c\textrm{-}b2$}

\addplot+[color=red, solid, mark=o, mark size=1.5pt, line width=0.6pt]
coordinates {
(75,52.71938158568603)
(80,51.427534471961)
(85,35.80213081220134)
(90,23.82227746627579)
(95,21.288257813243586)
(100,18.525291638480322)
};
\addlegendentry{$g_c\textrm{-}b3$}

\addplot+[color=green!50!black, solid, mark=o, mark size=1.5pt, line width=0.6pt]
coordinates {
(75,311.9843134326613)
(80,319.92395754179523)
(85,293.6185890098381)
(90,254.29779777085432)
(95,245.93128201747768)
(100,212.0861060504624)
};
\addlegendentry{$g_c\textrm{-}b27$}

\end{axis}
\end{tikzpicture}
\end{adjustbox}
\vspace{-0.9cm}
\caption{Average net profits of SGs at buses \{2,3,27\} over the horizon, calculated by the dispatchable pricing method. The profits of the remaining SGs are zero since they are not dispatched.}
\label{fig:dispatchable_net_profit}
\vspace{-0.5cm}
\end{figure}
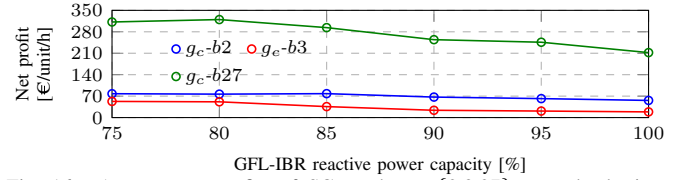

The ancillary service prices under different reactive power capacities of GFL-IBR are presented in Fig. \ref{fig:mean_prices_dispatchable}. Consistent with the pricing results of the restricted method in Fig.~\ref{fig:mean_prices_restricted}, the service price at bus 24 remains higher than that at bus 23 before the reactive power capacity reaches $95\%$, and is then surpassed by bus 23. In addition, the price of reactive power is still always higher than the SCR price for voltage stability maintenance.

Fig. \ref{fig:dispatchable_net_profit} illustrates the variation in net profits of market clearing units. As reactive power plays an increasingly important role in maintaining voltage stability, the demand for SCR diminishes gradually. Meanwhile, rising wind power penetration (as in Fig.~\ref{fig:wind_curt_and_cost}) curtails the output of thermal units. These factors jointly lead to a continuous decline in the net profits of these units. 


Overall, none of the units operate at a loss within this 24‑hour test instance, ensuring the financial sustainability of committed units for the considered case. This observation indicates that dispatchable pricing is capable of providing adequate incentives solely through market clearing prices under the studied system conditions.

\subsection{Impact of Offline Training Errors on Pricing}\label{Impact of Offline Training Errors on Pricing}

\begin{figure}[t]
\centering
\begin{adjustbox}{minipage=[t]{1\linewidth},center}
\begin{tikzpicture}
\begin{axis}[
    width=1\linewidth, height=3cm,
    xlabel={\scriptsize Time [h]},
    ylabel={\scriptsize \parbox[c]{3cm}{\centering Change of $\lambda_{g_c,\mathrm{commit}}$ \\[1pt]{[\texteuro/unit]}}},
    ylabel style={align=center, yshift=-0.3cm},
    label style={font=\scriptsize},
    tick label style={font=\scriptsize},
    xmin=1, xmax=24,
    ymin=-50, ymax=300,
    ytick distance=50,
    xtick={4,8,12,16,20},
    grid=both, grid style={dashed, gray!60},
    scaled y ticks = false,
    y tick label style={
        /pgf/number format/fixed,
        /pgf/number format/precision=0
    },
    legend style={
        font=\scriptsize,
        fill=none,
        draw=none,
        at={(0.173,0.18)}, anchor=south,
        legend columns=2
    },
    legend cell align={left},
    legend image post style={scale=1},
    legend image code/.code={%
        \draw[mark repeat=1,mark phase=1,#1] plot coordinates {(0,0)};
    },
    axis lines=box
]
\addplot+[  color=blue, solid, mark=star, mark size=1.5pt, line width=0.6pt]
coordinates {
(1,0.0)
(2,0.0)
(3,0.0)
(4,0.0)
(5,0.0)
(6,0.0)
(7,0.0)
(8,0.0893)
(9,236.2991)
(10,0.0)
(11,0.0)
(12,0.0)
(13,0.0)
(14,0.0)
(15,0.0)
(16,0.0)
(17,255.3208)
(18,0.0)
(19,228.9986)
(20,0.0)
(21,236.3800)
(22,0.0)
(23,0.0)
(24,0.0)
};
\addlegendentry{$g_c$-$b2$}

\addplot+[  color=red, solid, mark=o, mark size=1.5pt, line width=0.6pt]
coordinates {
(1,0.0)
(2,0.0)
(3,0.0)
(4,0.0)
(5,0.0)
(6,0.0)
(7,0.0)
(8,0.0)
(9,0.0)
(10,0.0)
(11,0.0)
(12,0.0)
(13,0.0)
(14,0.0)
(15,0.0)
(16,0.0)
(17,273.1383)
(18,0.0)
(19,245.0648)
(20,0.0)
(21,0.0)
(22,0.0)
(23,0.0)
(24,0.0)
};
\addlegendentry{$g_c$-$b3$}

\addplot+[  color=green!50!black, solid, mark=o, mark size=1.5pt, line width=0.6pt]
coordinates {
(1,0.017879056052265696)
(2,0.01844813219435082)
(3,0.018009930736525348)
(4,0.01777027023069877)
(5,0.018199792118934965)
(6,0.018517741645553087)
(7,-0.002526139222538859)
(8,-0.03808647947147392)
(9,261.52669116952467)
(10,197.93346151712035)
(11,197.9334469446302)
(12,197.93355645385515)
(13,197.93352464064753)
(14,197.93350096148885)
(15,197.9316144344555)
(16,197.93353678038744)
(17,255.32656358405248)
(18,197.96703205228903)
(19,261.5267307325)
(20,197.93362714492935)
(21,261.48877348290586)
(22,197.94530585153797)
(23,0.018154512094156416)
(24,0.01833108973482922)
};
\addlegendentry{$g_c$-$b27$}
\end{axis}
\end{tikzpicture}
\end{adjustbox}
\vspace{-1cm}
\caption{Differences of commitment prices captured by SGs at buses \{2, 3, 27\}, calculated by new values minus original values in Fig.~\ref{fig:UC_price}, in each hour.}
\label{fig:UC_price_variations}
\vspace{-0.5cm}
\end{figure}

Although Table~\ref{table:Approximation_Errors} has demonstrated the effectiveness of the offline-trained impedance terms presented in Appendix~\ref{Approximation of Nonlinear Impedance Ratios}, the approximation errors introduced by these terms would potentially affect the pricing results. It is relevant to investigate such impact to further verify the proposed pricing schemes. Therefore, the following validation procedure is designed:

\begin{adjustwidth}{0.5cm}{0.5cm}
\textbf{Step 1:} Run the UC with approximated voltage stability constraints to confirm optimal commitment decisions `$u_{g_c}^*$' and calculate the actual impedance terms `$\mathbf{z}_{g_f}^{g_f^\prime}, \mathbf{z}_{g_f}^1$' in the voltage stability constraints for each time period.

\noindent \textbf{Step 2:} Set the actual terms as the benchmark and calculate the MAPE between the shadow prices associated with the actual and approximate constraints.
\end{adjustwidth}

It should be noted that this validation is conducted only for the restricted pricing. This is because the marginal unit pricing has been demonstrated to be unsuitable for pricing, whereas the dispatchable pricing relaxes the commitment variables and may therefore fail to accurately reflect the variations in system impedances, potentially magnifying or attenuating the approximation errors. Moreover, fixed costs would be incorporated into the service prices, which is hard to decouple.

Taking the scheduling horizon of the case study as an example, the service prices in the two cases are the same, indicating that the values in Fig.~\ref{fig:Shadow prices of service_restricted} and Fig.~\ref{fig:hourly_energy_price} remain unchanged when the actual impedance terms are used. However, the solutions to the dual problems in the two cases are not exactly identical, as the objective values under the approximate and actual constraints are $27.01$ k\texteuro~and $26.94$ k\texteuro, respectively, corresponding to a difference of $0.26\%$. Further examination of the commitment prices reveals that $\lambda_{g_c,\mathrm{commit}}$ for SGs at buses~\{2, 3, 27\} obviously change when the actual impedance terms are used, as shown in Fig.~\ref{fig:UC_price_variations}. 

Clearly, the commitment prices obtained with the actual impedance terms increase, with the corresponding MAPEs being $\{10.44\%, 2.40\%, 58.34\%\}$, respectively. These errors are of a different order of magnitude from the MAPEs of the approximate impedance terms reported in Table~\ref{table:Approximation_Errors}. This result demonstrates that, although the offline training can accurately capture variations in the impedance, the dual problem is sensitive to the approximation errors and can reflect them in the resulting shadow prices.

Nevertheless, these pricing differences would not raise concerns. On the one hand, the energy and ancillary service prices remain unchanged, and the resulting price signals therefore remain efficient. On the other hand, all SGs can still obtain sufficient revenues with raised commitment prices to sustain their operation and stay in the market.

\subsection{Absence of Network Representation}\label{Absence of Network Representation}
It is worth noting that the proposed UC formulation uses single‑node power balance constraints \eqref{eq:cons_balance_active} and \eqref{eq:cons_balance_reactive}, without explicit line flow constraints, producing a uniform energy price across the system.

This model design choice is driven by the primary objective: to compare the economic characteristics of three non‑convex pricing schemes for voltage stability services. Introducing full Alternating Current (AC) power flow constraints into the UC model creates a problem which is NP‑hard and generally a large‑scale non‑convex nonlinear program \cite{shao2023linear}. Congestion‑induced price distortions would then obscure the core comparison of pricing mechanisms driven by commitment decisions and SOC‑constraint dual decomposition. 

In addition, topology features are not entirely discarded: pre‑trained impedance‑ratio coefficients in Appendix~\ref{Approximation of Nonlinear Impedance Ratios}, calculated from the full admittance matrix of the test system, encode electrical coupling and local SCR conditions for each GFL‑IBR bus. As a result, the shadow prices for SCR enhancement and reactive power support are bus‑specific, reflecting the heterogeneous marginal value of voltage stability services at different inverter‑connection nodes. The obtained bus‑dependent ancillary service prices provide useful locational reference signals indicating the marginal economic value of voltage stability support at each IBR bus and can support planning‑level analysis.

\subsection{Simplification on Physical Constraints of Thermal Units}\label{Simplification on Physical Constraints of Thermal Units}
Note that this MISOCP \eqref{eq:MISOCP} accounts for fixed costs, including no-load costs and startup/shutdown costs. Ramp-rate limits and minimum up/down time constraints are excluded, following the treatment in \cite{hytowitz2020impacts}, since incorporating these factors will affect pricing outcomes and potentially hinder an intuitive assessment of whether the shadow prices derived from the proposed pricing methods can sustain optimal system scheduling. Nevertheless, these constraints also constitute important sources of non‑convexity and uplift requirements in practical markets. Therefore, the obtained results are valid under this simplifying assumption, and care should be taken when directly extrapolating to real‑world system operation.

If ramp-rate and minimum up/down time constraints are incorporated, the UC model remains an MISOCP, since these constraints are linear. Under the restricted pricing, once the commitment states are fixed, the ramp-rate constraints become linear inequalities, and their associated dual variables consequently appear as additional components of the energy price. Although the minimum up/down time constraints do not introduce additional shadow prices, they would reshape the optimal commitment set and can therefore affect the final pricing outcomes. For example, the uplift payments may increase if SGs are forced to remain online due to minimum up-time limits. Therefore, the restricted pricing can be naturally extended to this case.

Under dispatchable pricing, the dual variables from these constraints would also introduce extra components into the shadow prices. However, relaxing the binary variables may distort the logical relationships imposed by these constraints, potentially amplifying the spurious price signals discussed in Section~\ref{Market Outcomes under Dispatchable Pricing Method} and further complicating the decoupling of fixed costs. Therefore, the treatment of these constraints under relaxation-based pricing needs further attention in future work.

\subsection{System Scalability and Computational Performance}

\begin{table}[t]
\centering
\caption{Time for Offline Training and Pricing Calculations under Different System Scales (seconds/test)}
\vspace{-0.3cm}
\setlength{\tabcolsep}{4pt}
{\fontsize{7.5pt}{10pt}\selectfont
\begin{tabular}{lcc}
\toprule
\diagbox{Test items}{Test systems}
 & \makecell{Modified IEEE\\30‑bus system}
 & \makecell{Modified IEEE\\57‑bus system} \\
\midrule
Offline training & $\mathrm{6.98}$  & $\mathrm{20.67}$ \\
Restricted pricing  & $\mathrm{12.54}$  & $\mathrm{21.74}$  \\
Dispatchable pricing & $\mathrm{2.91}$  & $\mathrm{3.05}$ \\
\bottomrule
\end{tabular}
}
\label{tab:computational_performance}
\vspace{-0.3cm}
\end{table}

The scalability of the proposed method is demonstrated through a modified IEEE 57-bus system \cite{pstca}, with the wind generation added at buses \{19, 23, 24\}. For this system, the MAPE from the offline training is up to $3.22\%$, which remains below the threshold. The computational times for the associated optimizations are summarized in Table~\ref{tab:computational_performance}.

The offline training procedure is a one‑time pre‑processing step performed prior to day‑ahead scheduling, and is not part of online dispatch computation. Its runtime increases from $6.98~s$ for the 30‑bus system to $20.67~s$ for the 57‑bus system, which remains manageable for pre‑computation.

The restricted pricing solves the MISOCP in \eqref{eq:MISOCP}, and its computation time increases from $12.54~s$ for the 30-bus system to $21.74~s$ for the 57-bus system. Despite this increase, the runtime remains acceptable for day-ahead UC applications, which generally have less stringent latency requirements. In contrast, dispatchable pricing relaxes the binary commitment variables and reduces the problem to an SOC program, resulting in a lower computation time. Moreover, its computation time increases only marginally, from $2.91~s$ to $3.05~s$, demonstrating favorable computational scalability.

Overall, both the offline training and pricing processes exhibit good scalability, with computation times increasing as the system size grows. Although the offline training is time-consuming, it only needs to be performed once for a given system and can then be applied to all operating conditions. Therefore, the pricing process is of greater practical concern, although its computation time can be further reduced with more powerful computational resources.

\section{Conclusion and Future Work}\label{Conclusion}
This paper proposes an SOC constraint-based shadow pricing formulation for static voltage stability services provided by thermal generators and RES. The scheme incentivizes generators to participate in the ancillary service market. Specifically, (V)SGs can be remunerated for improving SCR of weak buses where GFL-IBR are typically connected, while GFL-IBR can adjust their reactive power injection to support voltage in order to capture these rewards. This way, a cost-effective utilization of RES can be achieved without violating the voltage stability. 

To determine the economic value of ancillary services while addressing non-convexities, though intuitive, marginal unit pricing cannot compute shadow prices reflecting cross-service coupling and unit marginal system contributions, leaving market settlement undetermined and rendering the method inappropriate for pricing. However, it is demonstrated in the adopted case that duality-based restricted and dispatchable pricing methods can produce revenue-adequate shadow prices. Restricted pricing sets commitment prices to cover part of uplift payments for units and fully recovers their operating costs, which however necessitates well‑defined market rules for uplift cost allocation. By contrast, dispatchable pricing relaxes binary commitment constraints to embed fixed costs into market prices, yet comes at the cost of less efficient price signals that may deviate from true physical marginal contributions.

For practical market deployment, future work would incorporate full network representation into voltage stability service pricing. Future relevant areas also include considering additional stability services, such as frequency reserves and inertia, to design a market framework that fully leverages IBR capabilities, in order to support their role as dominant assets in future decarbonized grids. Furthermore, while preserving model integrality, it is worthwhile to investigate methods for appropriately pricing voltage stability services, either with dedicated mechanisms for allocating uplift payments or without the need for uplift payments. This is very relevant to relaxation-based pricing, as it does not inherently guarantee cost recovery and may still require uplift payments under more diverse operating scenarios. Finally, given that market participants are generally self-interested and thus bid strategically to maximize their own profits, the resulting impacts on market clearing prices are worthy of investigation.

\vspace{-0.2cm}
\appendices
\section{Approximation of Nonlinear Impedance Ratios}\label{Approximation of Nonlinear Impedance Ratios}
The actual values obtained from the admittance matrix inversion are expressed as $\mathbf{z}_{g_f}^{g_f^\prime}=|Z_{\Phi({g_f})\Phi(g_f^\prime)}|/|Z_{\Phi({g_f})\Phi({g_f})}|$ and $\mathbf{z}_{g_f}^1=1/|Z_{\Phi({g_f})\Phi({g_f})}|$. The corresponding approximate values, denoted by $z_{g_f}^{g_f^\prime}$ and $z_{g_f}^1$, are determined as follows:
\begin{subequations}\label{eq:approx_z_delta}
\begin{align}
z_{g_f}^{\delta,\mathrm{I}} &= 
\sum_{g_c \in \mathcal{G}_c} k_{{g_f},g_c}^{\delta,\mathrm{I}} u_{g_c}
+ \sum_{g_v \in \mathcal{G}_v} k_{{g_f},g_v}^{\delta,\mathrm{I}} \alpha_{g_v} \nonumber\\
&\quad + \sum_{m \in \mathcal{M}} k_{{g_f},m}^{\delta,\mathrm{I}} \eta_m,
\quad~~ \text{if } error\text{-}\mathrm{I} \leq error\text{-}\mathrm{II} \label{eq:approx_z_delta_I} \\
z_{g_f}^{\delta,\mathrm{II}} &= 
k_{g_f}^{\delta,\mathrm{II}} -(
\sum_{g_c \in \mathcal{G}_c} k_{{g_f},g_c}^{\delta,\mathrm{II}} u_{g_c}
+ \sum_{g_v \in \mathcal{G}_v} k_{{g_f},g_v}^{\delta,\mathrm{II}} \alpha_{g_v}  \nonumber\\
&\quad + \sum_{m \in \mathcal{M}} k_{{g_f},m}^{\delta,\mathrm{II}} \eta_m ),
~~~~ \text{if } error\text{-}\mathrm{I} \ge error\text{-}\mathrm{II} \label{eq:approx_z_delta_II} \\
& error\textrm{-}{\{ \mathrm{I},\mathrm{II} \}}: \quad \min _{\mathcal{K}^{\delta}} \sum_{\omega \in \Omega}\left(z_{g_f}^{\delta,{\{ \mathrm{I},\mathrm{II} \}}(\omega)}-\mathbf{z}_{{g_f}}^{\delta(\omega)}\right)^{2} \label{eq:approx_error_I_II}
\end{align}
\end{subequations}
where $\delta\in\{g_f^{\prime},1\}$ is introduced as an index to simplify notation. $\mathcal{K}^{\delta}=\{k_{g_f}^{\delta}, k_{{g_f},g_c}^\delta, k_{{g_f},g_v}^\delta, k_{{g_f},m}^\delta\}$ denotes the set of approximation coefficients. The VSG capacity percentage, $\alpha_{g_v}$ is a historical value. $\omega=\{u_{g_c}^{(\omega)},\alpha_{g_v}^{(\omega)},\mathbf{z}_{g_f}^{\delta(\omega)}\}\in \Omega$ denotes the dataset constructed by enumerating all operating states of (V)SGs. The term $k_{g_f,m}^\delta\eta_m, m\in \mathcal{G}_c \bigcup \mathcal{G}_v$ describes interactions between pairs of (V)SGs through the product of their operating states. The selection of approximation form \eqref{eq:approx_z_delta_I} or \eqref{eq:approx_z_delta_II} depends on the two error measures defined in \eqref{eq:approx_error_I_II}, which are obtained from a minimization problem used to fit the approximated values to the actual ones. An example implementing this procedure is given in \cite{Code}.

It is worth noting that the offline training is performed prior to system dispatch under fixed network topology parameters. All candidate commitment states of SGs are enumerated within the training dataset, so that the resultant fitting coefficients can fully account for the varying operating conditions driven by UC decisions. As long as the resulting approximation errors remain within the acceptable threshold, the robustness is sufficient to capture the variations in system impedance; and no retraining is required for this fixed network topology.

\vspace{-0.3cm}
\section{Procedure of Marginal Unit Pricing}\label{Procedure of Marginal Unit Pricing}
First, one must solve modified versions of the original MISOCP in \eqref{eq:MISOCP}, where the contribution of each generator to the voltage stability is sequentially removed. This model is given by the following compact form:
\begin{subequations}
    \begin{align}
   & \displaystyle \min  \quad f(\textbf{x}) \label{eq:marginal_unit_obj} \\
    & \mathrm{s.t.} ~\hspace{0.2cm} \hat{P}_{{g_f}}^{2}+\hat{Q}_{g_f}^{2} \leq \big( \hat{Q}_{g_f} + \Gamma_{g_f} \big)^{2} \label{eq:marginal_unit_soc_con}  \\ 
    & \qquad \hat{Q}_{g_f} \hspace{-0.1cm} = \hspace{-0.1cm} \Delta Q_{g_f} \hspace{-0.1cm} + \hspace{-0.8cm} \sum_{\substack{{g_f}' \in \mathcal{G}_f, {g_f}' \neq {g_f}}} \hspace{-0.7cm} z_{{g_f}}^{{g_f}'} \Delta Q_{{g_f}'} \hspace{-1pt}  ~| ~\{ z_{{g_f}}^{{g_f}'} | \mathcal{K}^{\delta} + \Delta \mathcal{K}^{\delta}      \} \\ 
   &  \qquad \Gamma_{g_f} = \frac{1}{2}z_{{g_f}}^{1} ~| ~\{ z_{{g_f}}^{1} | \mathcal{K}^{\delta} + \Delta \mathcal{K}^{\delta}      \} \label{eq:marginal_unit_gamma_con} \\
   & \qquad g(\textbf{x})\leq 0 \label{eq:marginal_unit_ineq_con}  \\
   &  \qquad h(\textbf{x})=0 \label{eq:marginal_unit_eq_con} \\
   &
\begin{bmatrix}
\Delta Q_{g_f} \\
\Delta \mathcal{K}^{\delta}
\end{bmatrix}^{\mathsf{T}}
\hspace{-0.25cm} =
\begin{cases}
\left[  0,\ -\mathcal{K}^{\delta} \right], \text{evaluate prices for a given unit} \\
\left[  Q_{g_f},\ 0 \right], \text{otherwise} 
\end{cases}
    \end{align}
\end{subequations}
where `$\textbf{x}$' represents the vector of decision variables and $f(\textbf{x})$ the system cost. Eqs.~\eqref{eq:marginal_unit_soc_con}-\eqref{eq:marginal_unit_gamma_con} are the variable voltage stability constraints. Eqs.~\eqref{eq:marginal_unit_ineq_con}-\eqref{eq:marginal_unit_eq_con} indicate all other constraints in \eqref{eq:MISOCP}, which remain unchanged. To determine the economic value of the voltage stability support provided by an IBR $g_f$, its contribution to equivalent reactive power injections needs to be removed, that is, $ \Delta Q_{g_f} $ is set as $0$. While $-\mathcal{K}^{\delta}$ is taken for evaluating the ancillary service prices for (V)SGs. 

\balance


\begin{thebibliography}{25}

\bibitem{chu2022voltage}
Z.~Chu and F.~Teng, “Voltage stability constrained unit commitment in power systems with high penetration of inverter-based generators,” \textit{IEEE Transactions on Power Systems}, vol.~38, no.~2, pp.~1572–1582, 2022.

\bibitem{hosseinzadeh2021voltage}
N.~Hosseinzadeh, A.~Aziz, A.~Mahmud, A.~Gargoom and M.~Rabbani, “Voltage stability of power systems with renewable-energy inverter-based generators: A review,” \textit{Electronics}, vol.~10, no.~2, pp.~115, 2021.

\bibitem{qays2023system}
M.~O.~Qays, I.~Ahmad, D.~Habibi, A.~Aziz and T.~Mahmoud, “System strength shortfall challenges for renewable energy-based power systems: A review,” \textit{Renewable and Sustainable Energy Reviews}, vol.~183, pp.~113447, 2023.

\bibitem{savvopoulos2019long}
N.~Savvopoulos, C.~Y.~Evrenosoglu, A.~Marinakis, A.~Oudalov and N.~Hatziargyriou, “A long-term reactive power planning framework for transmission grids with high shares of variable renewable generation,” in \textit{2019 IEEE Milan PowerTech}, pp.~1–6, 2019.

\bibitem{roy2013reactive}
N.~K.~Roy, H.~R.~Pota and M.~J.~Hossain, “Reactive power management of distribution networks with wind generation for improving voltage stability,” \textit{Renewable Energy}, vol.~58, pp.~85–94, 2013.

\bibitem{wu2025enhancing}
K.~Wu, J.~Hao, Z.~Chen, J.~You, S.~Cao, W.~Tang and X.~Zhu, “Enhancing grid strength in high-renewable systems: Selecting retired thermal power units retrofit to synchronous condensers based on multi-dimensional evaluation method,” \textit{Electronics}, vol.~14, no.~12, pp.~2467, 2025.

\bibitem{guo2022control}
Z.~Guo, X.~Zhang, M.~Li, H.~Wang, F.~Han, X.~Fu and J.~Wang, “Control and capacity planning for energy storage systems to enhance the stability of renewable generation under weak grids,” \textit{IET Renewable Power Generation}, vol.~16, no.~4, pp.~761–780, 2022.

\bibitem{khan2024grid}
M.~Khan, W.~Wu and L.~Li, “Grid-forming control for inverter-based resources in power systems: A review on its operation, system stability, and prospective,” \textit{IET Renewable Power Generation}, vol.~18, no.~6, pp.~887–907, 2024.

\bibitem{kim2025seamless}
H.~J.~Kim, H.~J.~Lee, M.~S.~Kim and E.~S.~Lee, “Seamless transition control method of grid-connected inverters under unbalanced voltage conditions,” \textit{IEEE Open Journal of Power Electronics}, 2025.

\bibitem{mandoulidis2022overview}
P.~Mandoulidis, G.~Chaspierre, C.~Vournas and T.~Van Cutsem, “Overview, comparison, and extension of emergency controls against voltage instability using Inverter-Based Generators,” \textit{Sustainable Energy, Grids and Networks}, vol.~31, pp.~100710, 2022.

\bibitem{du2026improved}
H.~Du, W.~Guo, J.~Lan, Z.~He and W.~Hou, “Improved adaptive virtual impedance control for virtual synchronous generators based on sequence impedance stability analysis,” \textit{Electric Power Systems Research}, vol.~254, pp.~112636, 2026.

\bibitem{banshwar2017renewable}
A.~Banshwar, N.~K.~Sharma, Y.~R.~Sood and R.~Shrivastava, “Renewable energy sources as a new participant in ancillary service markets,” \textit{Energy Strategy Reviews}, vol.~18, pp.~106–120, 2017.

\bibitem{chattopadhyay2003spot}
D.~Chattopadhyay, B.~B.~Chakrabarti and E.~G.~Read, “A spot pricing mechanism for voltage stability,” \textit{International Journal of Electrical Power \& Energy Systems}, vol.~25, no.~9, pp.~725–734, 2003.

\bibitem{kim2011market}
M.~K.~Kim, J.~K.~Park and Y.~W.~Nam, “Market-clearing for pricing system security based on voltage stability criteria,” \textit{Energy}, vol.~36, no.~2, pp.~1255–1264, 2011.

\bibitem{potter2023reactive}
A.~Potter, R.~Haider, G.~Ferro, M.~Robba and A.~M.~Annaswamy, “A reactive power market for the future grid,” \textit{Advances in Applied Energy}, vol.~9, pp.~100114, 2023.

\bibitem{chung2004cost}
C.~Y.~Chung, T.~S.~Chung, C.~W.~Yu and X.~J.~Lin, “Cost-based reactive power pricing with voltage security consideration in restructured power systems,” \textit{Electric Power Systems Research}, vol.~70, no.~2, pp.~85–91, 2004.

\bibitem{gu2019review}
H.~Gu, R.~Yan and T.~Saha, “Review of system strength and inertia requirements for the national electricity market of Australia,” \textit{CSEE Journal of Power and Energy Systems}, vol.~5, no.~3, pp.~295–305, 2019.

\bibitem{gribik2007market}
P.~R.~Gribik, W.~W.~Hogan and S.~L.~Pope, “Market-clearing electricity prices and energy uplift,” Cambridge, MA, pp.~1–46, 2007.

\bibitem{chu2024pricing}
Z.~Chu, J.~Wu and F.~Teng, “Pricing of short circuit current in high IBR-penetrated system,” \textit{Electric Power Systems Research}, vol.~235, p.~110690, 2024.

\bibitem{taylor2015convex}
J.~A.~Taylor, \textit{Convex optimization of power systems}, Cambridge University Press, 2015.

\bibitem{ruiz2012pricing}
C.~Ruiz, A.~J.~Conejo and S.~A.~Gabriel, “Pricing non-convexities in an electricity pool,” \textit{IEEE Transactions on Power Systems}, vol.~27, no.~3, pp.~1334–1342, 2012.

\bibitem{badesa2022assigning}
L.~Badesa, C.~Matamala, Y.~Zhou and G.~Strbac, “Assigning shadow prices to synthetic inertia and frequency response reserves from renewable energy sources,” \textit{IEEE Transactions on Sustainable Energy}, vol.~14, no.~1, pp.~12–26, 2022.

\bibitem{hytowitz2020impacts}
R.~B.~Hytowitz, B.~Frew, G.~Stephen, E.~Ela, N.~Singhal, A.~Bloom and J.~Lau, “Impacts of price formation efforts considering high renewable penetration levels and system resource adequacy targets,” National Renewable Energy Laboratory (NREL), Golden, CO, 2020.

\bibitem{matamala2024cost}
C.~Matamala, L.~Badesa, R.~Moreno and G.~Strbac, ``Cost allocation for inertia and frequency response ancillary services,'' \textit{IEEE Transactions on Energy Markets, Policy and Regulation}, vol.~2, no.~3, pp.~328--338, 2024.


\bibitem{Code}
P.~Wang, ``Repository of Pricing Voltage Stability Services'', 2026. \url{https://github.com/pwang30/Pricing_voltage_stability_services}

\bibitem{shao2023linear}
Z.~Shao, Q.~Zhai, Z.~Han and X.~Guan, ``A linear AC unit commitment formulation: An application of data‑driven linear power flow model,'' \textit{International Journal of Electrical Power \& Energy Systems}, vol.~145, pp.~108673, 2023.


\bibitem{pstca}
``Power systems test case archive,'' Accessed: Sep. 21, 2026. [Online]. Available: \url{https://labs.ece.uw.edu/pstca/}

\end{thebibliography}
\end{document}